\documentclass[fleqn,usenatbib]{mnras}

\usepackage{newtxtext,newtxmath}
\usepackage{xspace}

\usepackage[T1]{fontenc}
\usepackage{ae,aecompl}
\usepackage{url}

\usepackage{graphicx}	
\usepackage{amsmath}	
\usepackage{amssymb}	
\usepackage{float}

\usepackage[utf8]{inputenc}

\title[3FHL TeV candidate spectroscopy]{Optical spectroscopy of BL Lac objects: TeV candidates \\}

\author[S. Paiano et al.]{Simona Paiano$^{1,2,3}$\thanks{E-mail:
simona.paiano@inaf.it}, Renato Falomo$^{4}$, Aldo Treves$^{3,5}$, Riccardo Scarpa$^{6,7}$\\
$^{1}$INAF - Osservatorio Astronomico di Roma, via Frascati 33, I-00040, Monteporzio Catone, Italy \\
$^{2}$INAF - IASF Milano, via Corti 12, I-20133, Milano, Italy \\
$^{3}$Universita' dell'Insubria, via Valeggio, 22100, Como, Italy\\
$^{4}$INAF - Osservatorio Astronomico di Padova, vicolo dell'Osservatorio 5, I-35122, Padova, Italy\\
$^{5}$INAF - Osservatorio Astronomico di Brera, via Bianchi 46, I-23807, Merate (Lecco), Italy\\
$^{6}$Instituto de Astrofisica de Canarias, C/O Via Lactea, s/n E38205 - La Laguna (Tenerife) - SPAIN\\
$^{7}$Universidad de La Laguna, Dpto. Astrofsica, s/n E-38206 La Laguna (Tenerife) - SPAIN
}

\date{Received:~\today; Accepted:~ }

\begin{document}
\label{firstpage}
\pagerange{\pageref{firstpage}--\pageref{lastpage}}
\maketitle

\begin{abstract}
We investigate the spectroscopic optical properties of gamma-ray sources detected with high significance above 50~GeV in the Third Catalog of Hard Fermi-LAT Sources (3FHL) and that are good candidates as TeV emitters.  
We focus on the 91 sources that are labelled by the \textit{Fermi} team as BL Lac objects (BLL) or Blazar candidates of uncertain type (BCU) , that are in the Northern hemisphere, and are with unknown or uncertain redshift. 
We report here on GTC spectra (in the spectral range 4100~-~7750~$\textrm{\AA}$) of 13 BCU and 42 BLL. 
We are able to classify the observed targets as BLL and each source is briefly discussed.
The spectra allowed us to determine the redshift of 25 objects on the basis of emission and/or absorption lines, finding 0.05~$<$~\textit{z}~$<$~0.91.
Most of the emission lines detected are due to forbidden transition of [O~III] and [N~II]. 
The observed line luminosity is found lower than that of QSOs at similar continuum and could be reconciled with the line-continuum luminosity relationship of QSOs if a significant beaming factor is assumed.
Moreover for 5 sources we found intervening absorption lines that allows to set a spectroscopic lower limit of the redshift. 
For the remaining 25 sources, for which the spectra are lineless, a lower limit to \textit{z} is given, assuming that the host galaxies are giant ellipticals.
\end{abstract}


\section{Introduction}  \label{sec:intro} 
BL Lac objects (BLL) are active galactic nuclei, whose emission is dominated by non-thermal radiation from a relativistically beamed jet aligned with the line of sight. The emission, of synchrotron-inverse Compton origin, extends over the whole electromagnetic spectrum,  from radio to gamma-ray.
Distinctive characteristics of the class are the large and rapid flux variability, high polarization, and the absence or weakness of spectral lines, washed out by the emission from the jet, that in most cases makes very arduous the determination of the their distance. 

Data-sets of BLL have been produced through a number of dedicated surveys in all spectral bands, originally in the radio \citep{stickel1991, giommi2002, marcha2013}, and X-rays \citep{stocke1990, perlman1996, perlman1998, lau1998, lau1999, voges1999, cusumano2010}, then in the optical and infrared \citep[e.g.][]{scarpa2000hubb, scarpa2000ir, urry2000, falomo2000, plotkin2010, dabrusco2014, dabrusco2019}, and more recently also in the gamma-ray band \citep[][and references in therein]{fermi20204fgl}. 
The latter surveys indeed showed that BLL represent the dominant extragalactic population of gamma sources \citep{fermi4lac2019}. 
Thanks to its systematic monitoring of the whole sky, and the long duration of the mission, the \textit{Fermi} catalogues of $\gamma$-ray sources have become a privileged means for discovering new BLL. 
The procedure of identification of the counterpart of the gamma-ray emitter generally goes through the search of X-ray and 
radio sources in the gamma-ray error box \citep[e.g.][]{stephen2010, takeuchi2013, landi2015, paiano2017sed}.
The BLL nature of the target can be guessed because of the broad spectral energy distribution (SED), which in BLL is well characterized by two broad humps, one due to synchrotron radiation (IR-optical), and one to inverse Compton (X-ray- Gamma rays). \citep[see e.g.][]{maraschi1992, ghisellini2009, ghisellini2017, madej2016}.
However, the secure classification derives generally from optical spectroscopy.
The identification of weak spectral lines is the basic tool for obtaining first of all a firm classification and the distance of the source, but also for constraining some physical parameters of the emission region. 
If no intrinsic lines are observed, there are indirect procedures based on some hypothesis on the nature of the host galaxy, which
can effectively produce lower limit to the redshift of the source \citep[e.g.][see also Section 4 of this paper]{sbarufatti2006a, 2006bsbarufatti, meisner2010, falomo2014}.
In some cases the optical spectra can exhibit absorption lines due to intervening medium, from which a spectroscopic lower limit to \textit{z} can be derived.

The knowledge of the distance is a basic ingredient for modeling the BLLs, for constructing the luminosity function of this class of blazars, to assess the cosmic evolution, and to evaluate the contribution of the BLL population to the gamma-ray cosmic background.
Because the overall spectrum may extend to the TeV domain, BLL of known redshift are unique probes of the extragalactic background light (EBL), since at z~$\geq$~1 the absorption effects due to pair production become important \citep{franceschini2017}. 
Some BLL are also supposedly ultrahigh energy (PeV) neutrino emitters \citep[][]{ahlers2015, padovani2016, righi2019}, and again the distance knowledge is a crucial information for modeling the neutrino production, and the cosmic neutrino background radiation.

All these arguments explain the major effort for optical spectroscopic studies of BLL candidates. 
Because of the above mentioned weakness of the lines, it is compulsory to secure spectra with high and adequate signal-to-noise (S/N) ratios, in order to be able to measure these features of very low equivalent widths (EW).
This requires the use of telescopes of the 8-10~m class.

For twenty years, we have embarked on a systematic study of BLL, lately focusing on \textit{Fermi} detections, using the ESO 8~m Very Large Telescope (VLT) and, more recently, the 10~m Gran Telescopio Canarias (GTC). 
Optical spectra of $\sim$250 BLL have been produced, adopting various selection criteria of the sources \citep{sbarufatti2005, sbarufatti2006a, sbarufatti2008, sbarufatti2009, landoni2012, landoni2013, landoni2014, landoni2015, paiano2016, falomo2017, paiano2017tev, paiano20173fgl, paiano2017ufo1, paiano2018txs, landoni2018, paiano2019ufo2}. 

Here we present optical spectroscopy of 55 objects, which exhibit a hard gamma-ray spectrum in the 3FHL catalogue \citep{ajello2017}. 
From these observations, we are able to characterize the optical spectral properties and we determine the redshift in many objects.

In a coordinated paper (Paiano et al, in preparation), we will point out candidates for detection in the TeV range on the basis of the extrapolation to the very high energy ($>$~$\sim$100~GeV) of the \textit{Fermi} spectra accounting for absorption by the EBL.  
It is expected that $\sim$20\% of the sources with known redshift considered in this work could be detected with the current Cherenkov telescopes, like MAGIC, and 85\% with the future Cherenkov Telescope Array (CTA).


\section{The sample} 
\label{sec:sample} 
The objects in our sample were extracted from the 3FHL catalog, which contains 1556 objects detected above $>$~10 GeV over 7 years of \textit{Fermi} operations. 
We concentrate on 246 \textit{TeV candidates}\footnote{flagged as "C" in the 3FHL catalog}, which are sources revealed with significance$>$3 at energy $>$~50 GeV. 
They exhibit a hard spectrum, with photon index $\Gamma\lesssim$~2.5 and integrated flux F(\footnotesize{$>$10~GeV})$\gtrsim$~10$^{-11}$ \textit{ph}~cm$^{-2}$ s$^{-1}$. 
Of these, 224 sources are classified by the \textit{Fermi} team as BLL and blazar candidates of uncertain type (BCU): 180 and 44 sources respectively. 
Extensive search in the literature show that 153 of these objects lack a measurement of the redshift or the available value is uncertain.
Of these, there are 21 BCU and 70 BLL at $\delta~>$-20 and thus well-observable in the Northern hemisphere. 

Here we report on extensive optical spectroscopy obtained at the GTC for 42 BLL and 13 BCU (see Table~\ref{tab:sample} for the properties of our sample). 

\section{Observations and data reduction} 
\label{sec:obsdata} 
Optical spectra were collected using the spectrograph OSIRIS \citep{cepa2003} of the 10.4~m GTC at the Roque de Los Muchachos (La Palma). 
We used the grism R1000B covering the spectral range 4100~-~7750~$\textrm{\AA}$, and a slit width~=~1.2”.

Data reduction was performed using the IRAF software \citep{tody1986, tody1993} and standard procedures for long slit spectroscopy, following the same scheme given in \citet{paiano2017tev, paiano20173fgl, paiano2019ufo2}.  
For each source at least three spectra were obtained and combined together to obtain the final spectrum.
Cosmic rays were removed using the L.A.Cosmic algorithm \citep{vandokkum2001}. 
The accuracy of the wavelength calibration is $\sim$0.1~$\textrm{\AA}$ over the whole observed spectral range. 
The relative flux calibration was performed from the observation of a number of spectro-photometric standard stars secured during each night of the program. 
Since most of the spectra were obtained during non-photometric nights, we set the absolute calibration from the comparison of the magnitude of the targets, as from the acquisition image in the g band, and that of secondary photometric objects in the field as derived from PanSTARRS or SDSS images. 
Finally all spectra were dereddened applying the extinction law by \citet{cardelli1989} and assuming the value of Galactic extinction \textit{E(B-V)} derived from the NASA/IPAC Infrared Science Archive 6\footnote{http://irsa.ipac.caltech.edu/applications/DUST/} \citep{schlafly2011}.
In Fig.~\ref{fig:spectra}, we display all our absolute-calibrated and dereddened spectra (\textit{upper panel}) with the normalized form (\textit{bottom panel}) obtained through the fit of the continuum with a cubic spline. 
The spectra are also available in our online database ZBLLAC\footnote{http://web.oapd.inaf.it/zbllac/}. 

\section{Methods and results} 
\label{sec:results}
For each spectrum the S/N was measured, and the minimum detectable EW was evaluated following the procedure described in detail in \citet{paiano2017tev}. 

The optical spectrum is assumed to be the superposition of a non thermal component described by a power law (PL)  and the starlight component of the host galaxy. The latter is in almost all cases a giant elliptical galaxy of M(R)$\sim$-22.9 \citep{sbarufatti2005imaging}.
We performed a spectral decomposition using these two components for all our targets (see some examples in Fig.~\ref{fig:decomposition}) and for details for the individual sources are reported in the corresponding notes. 
The visibility of the absorption features due to the host galaxy at a given wavelength, depends on the ratio of the two components (N/H, nucleus-to-host ratio) and on the minimum detectable EW \citep{sbarufatti2006a}. 
Under this assumption, it is possible to set a lower limit to the redshift, when no absorption lines of the host galaxy are detected. We followed the procedure described in the Appendix A of \citet{paiano2017tev} to evaluate this redshift lower limit.

Our optical spectroscopy shows that all 55 observed targets are BLL, including the 13 targets previously classified as BCU.
No objects like flat spectrum radio-quasar or narrow line Seyfert~I were found. 
In Tab~\ref{tab:results} we report for each target whether the spectrum exhibits emission lines (labelled as E), galactic lines (G), intervening lines (I) or if it is lineless (L). 
For each source we give the redshift, the spectroscopic redshift lower limit in case of intervening line detection or a lower limit, estimated following the minimum EW method (described above), when the spectrum is featureless. 
In Table~\ref{tab:ew}, we give the measurement of the all detected lines and in Fig.~\ref{fig:closeup} we display some close-ups of the normalized spectra around interesting detected features.
In 12 objects we detected emission lines and in 20 we found intrinsic absorption lines. This allows us the determination of 26 new redshifts, ranging from z~=~0.05 to z~=~0.91. The mean value is~$\sim$0.28.
In 5 sources intervening lines are detected, in most case due to Mg~II, which enable to set a robust spectroscopic lower limit to the redshift. 
These intervening lines are originated at redshift in the range between 0.1 and 1.1. 
The intervening lines are an efficient tool for finding the most distant sources.
For the reminder 24 sources, we provide the redshift lower limits based on lack of detection of spectral features. These limits range from z~$>$~0.1 to z~$>$~0.6.

\setcounter{figure}{0}
\begin{figure*}
\includegraphics[width=1.3\textwidth, angle=-90]{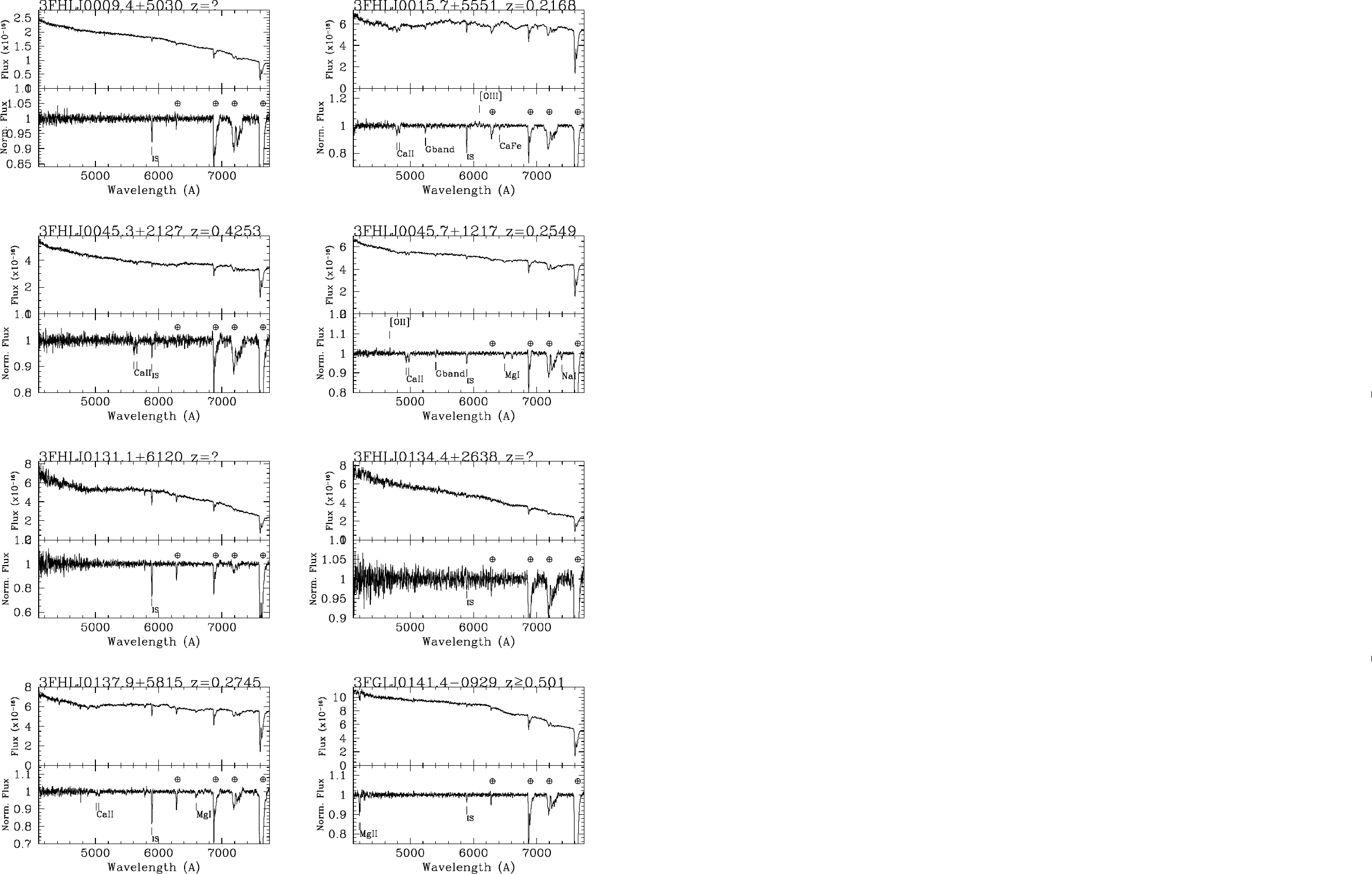}
\caption{Spectra of the UGSs obtained at GTC. \textit{Top panel}: Flux calibrated and dereddened spectra. \textit{Bottom panel}: Normalized spectra. The main telluric bands are indicated by $\oplus$, the absorption features from interstellar medium of our galaxies are labelled as IS (Inter-Stellar)}
\label{fig:spectra}
\end{figure*}

\setcounter{figure}{1}
\begin{figure*}
\centering
\includegraphics[width=0.95\textwidth]{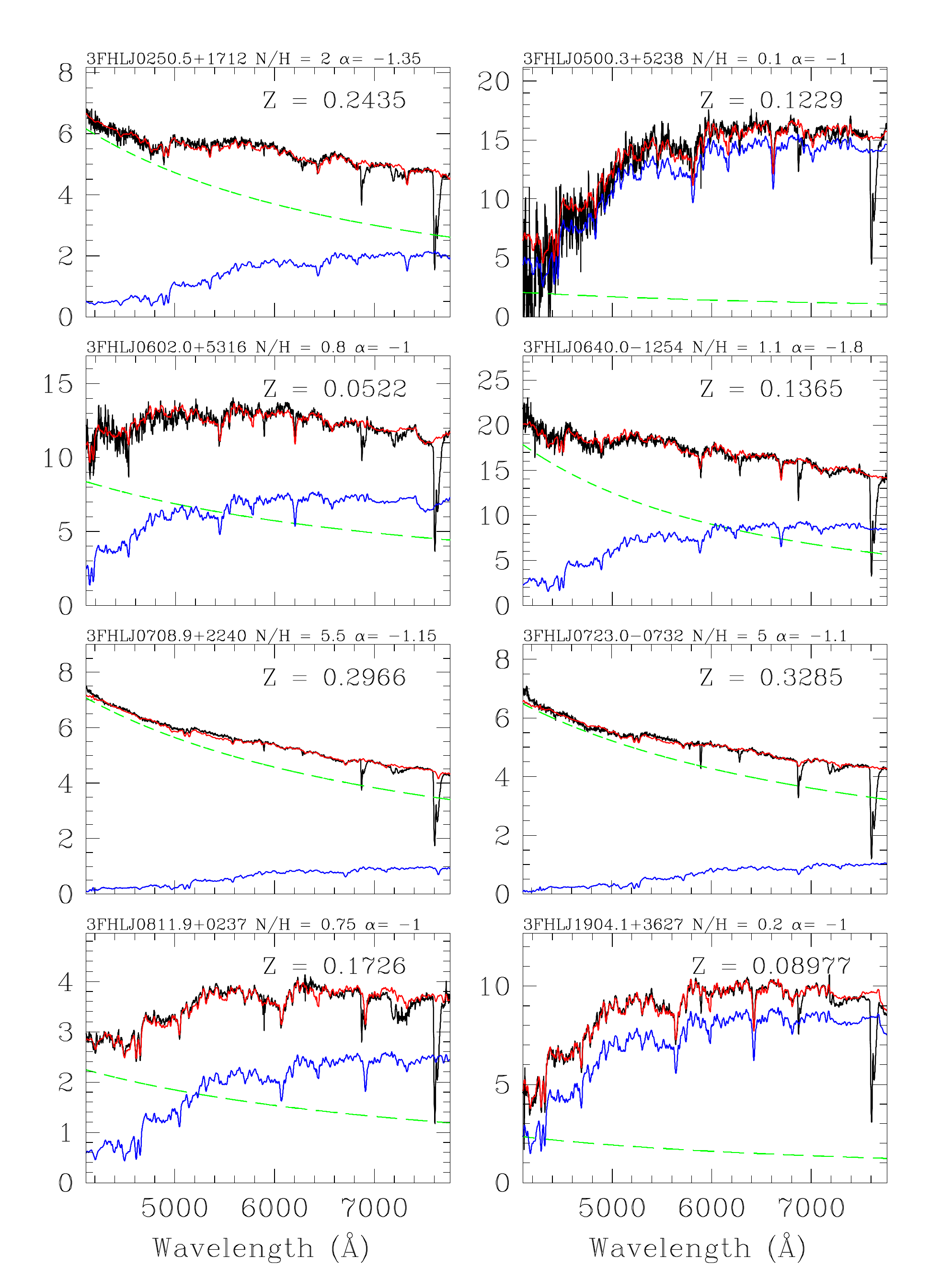}
\caption{Spectral decomposition of the observed optical spectrum (black line) of selected targets into a power law ( green dashed line ) and a template of elliptical for the host galaxy (blue line). The fit is given by the red solid line (see text for details). On each panel the nucleus to host ratio is given together with the spectral index of the power law component.}
\label{fig:decomposition}
\end{figure*}

\setcounter{figure}{2}
\begin{figure*}
\centering
\includegraphics[width=1.25\textwidth, angle=-90]{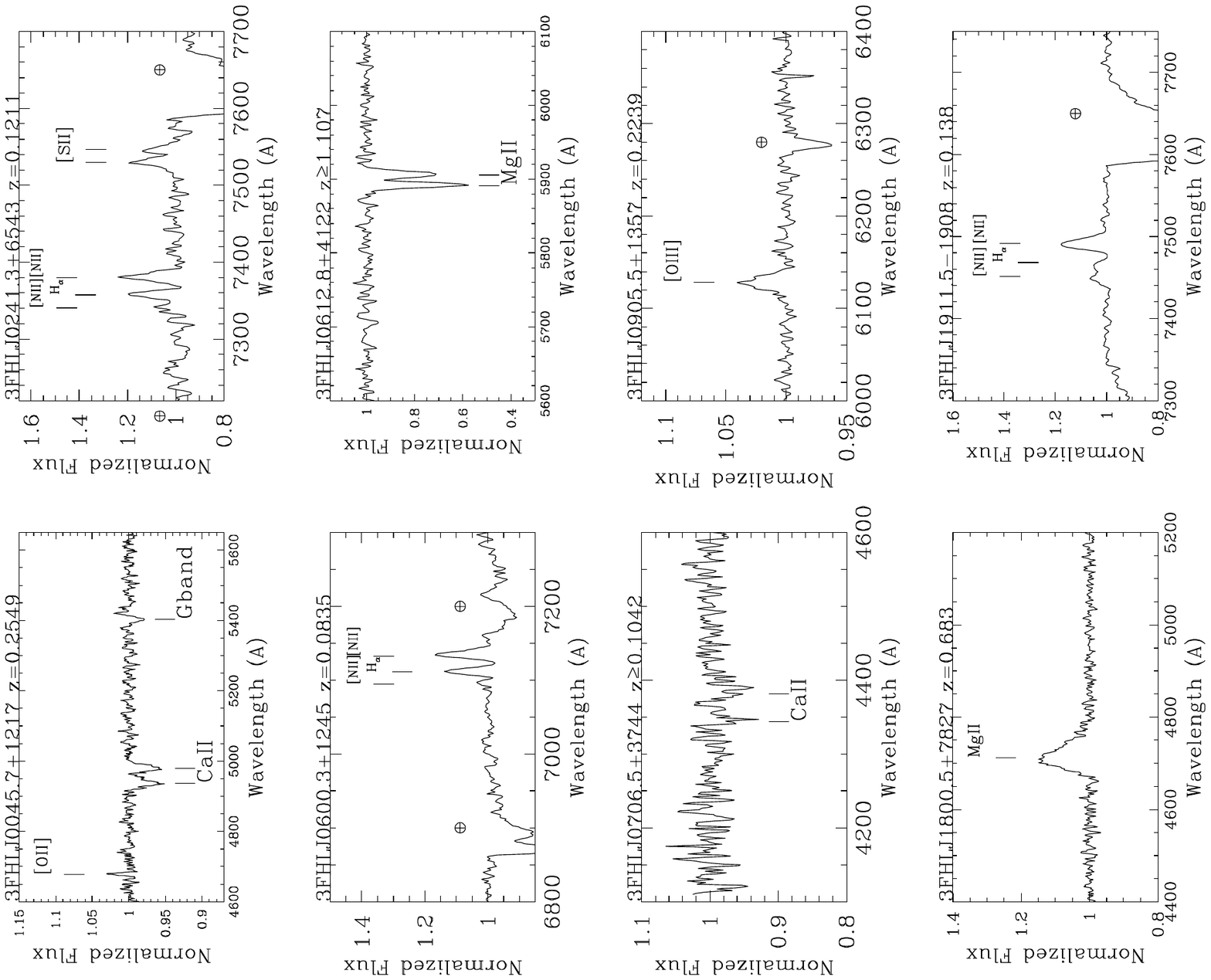}
\caption{Some examples of close-up of the normalized spectra, obtained at GTC of the 3FHL TeV candidate objects, around the detected spectral lines. 
Main telluric bands are indicated as $\oplus$, spectral lines are marked by line identification.}
\label{fig:closeup}
\end{figure*}


\section{Notes on individual sources} 
\label{sec:notes}
\begin{itemize}
\item[] \textbf{3FHL~J0009.4+5030}: %
The spectrum appears featureless and no emission/absorption lines are detected.
On the basis of the minimum EW ($\sim$0.20~$\textrm{\AA}$) and on the lack of detection of features from the host galaxy, we can set a redshift lower limit of z~$>$~0.6.

\item[] \textbf{3FHL~J0015.7+5551}: %
The observed optical counterpart (g~=~18.7) of this gamma-ray source is at (RA,DEC)=(00:15:40,~+55:51:45) and spatially coincident there is one X-ray source (J001540+555144), detected by \textit{Swift}, and one radio source (NVSS~J001540+555144).
The de-reddened (\textit{E(B-V)}~=~0.37) spectrum of the source is characterized by stellar absorption features of the host galaxy (Ca~II 3934,3968, G-band 4305, Ca+Fe 5269) at z~=~0.2168, superposed to a non thermal power law continuum (F$_{\lambda} \propto \lambda^{\alpha}$, $\alpha$~=~-1.2). 
From the spectral decomposition of the two components, we set a N/H~=~1.5.
This object was previously studied by \citet{alvarez2016a}, who due to low S/N of the spectrum failed to detect the absorption features.  
At $\sim$2.5" (West on the right), there is a faint companion (g~=~20.7) that was placed in the slit during our observation and exhibits a stellar spectrum. 
 
\item[] \textbf{3FHL~J0045.3+2127}: %
Our optical spectrum (S/N$\sim$90) is characterized by a power law emission ($\alpha$~=~-1.3). 
We detect an faint absorption doublet at $\sim$5600~$\textrm{\AA}$ that we identify as Ca~II 3934,3968 at z~=~0.4253.
It is worth to note that there is a galaxy close to our target at a projected distance of $\sim$10" (see Fig.~\ref{fig:0045imaging}) with redshift z~=~0.4265 provided by SDSS survey. 
A previous spectrum provided by SDSS survey failed to detect any feature, while \citet{shaw2013} provided a lower limit compatible with our measurement. 

\setcounter{figure}{3}
\begin{figure}
\centering
\includegraphics[width=0.35\textwidth]{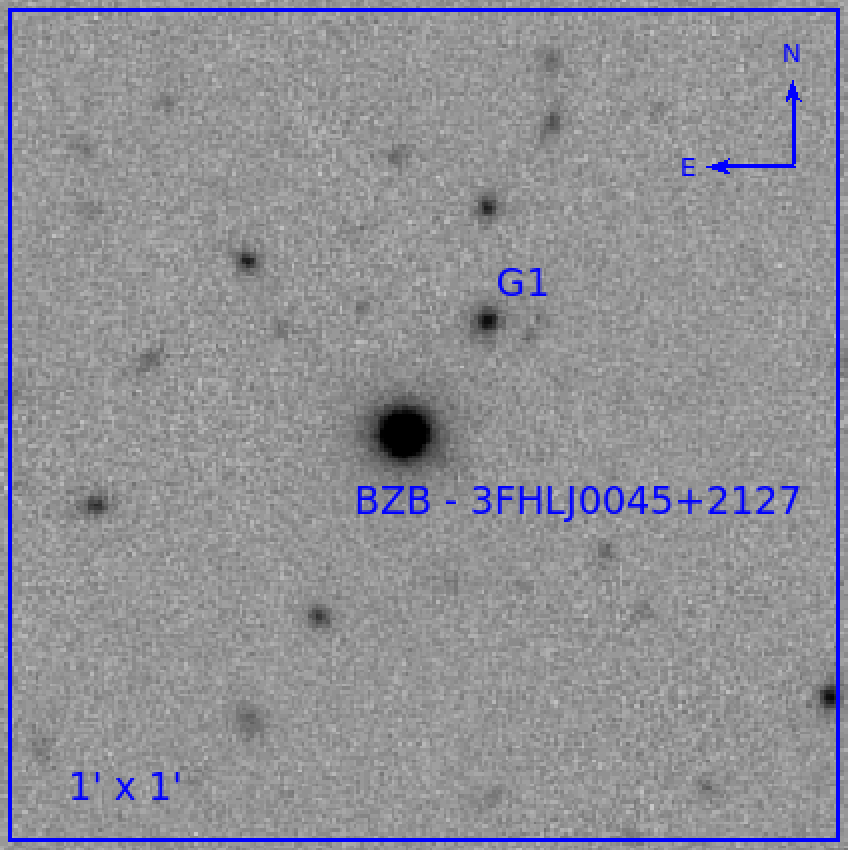}
\caption{PANSTARR \textit{i}-band image of the BLL 3FHL~J0045+2127. There is one close companion at $\sim$10$^{\prime\prime}$ from the target (that is at the center). Field shown is 1$^{\prime}$, North up and East left.}
\label{fig:0045imaging}
\end{figure}

\item[] \textbf{3FHL~J0045.7+1217}: %
In our spectrum we find absorption lines due to Ca~II 3934,3968, G-band~4305, Mg~I 5175 and Ca+Fe~5269 attributed to the stellar population of the host galaxy.
In addition we also detect a faint (EW~=~0.2~$\textrm{\AA}$) emission line due to [O~II]~3727 (see Fig.~\ref{fig:closeup}).
The redshift is 0.2549. 
This is consistent with the lower limit proposed by \citet{shaw2013}.
From the spectral decomposition, we set a N/H$\sim$4.5 and $\alpha$~=~-1.0 for the nuclear power law component.

\item[] \textbf{3FHL~J0131.1+6120}: %
The dereddened spectrum (\textit{E(B-V)}~=~0.79) is featureless with a minimum detectable EW varying between 0.20~$\textrm{\AA}$ and 0.70~$\textrm{\AA}$ along the spectrum. 
We can set a lower limit of the redshift z~$>$~0.10.

\item[] \textbf{3FHL~J0134.4+2638}: %
We obtain an optical spectrum that is featureless. 
No emission or absorption lines are detected out to a minimum EW of 0.35~-~0.50~$\textrm{\AA}$.
This is consistent with the featureless spectrum of the SDSS survey.
From the lack of detection of host galaxy features, we can set a redshift lower limit z~$>$~0.15.
An optical spectrum obtained by \citet{marchesi2018} exhibits a prominent emission line at $\sim$~4400~$\textrm{\AA}$ which is interpreted as Mg~II~2800. 
We suspect that the reported line is a spurious artifact. In fact this feature is not reported in the SDSS spectrum and in \citet{shaw2013}.

\item[] \textbf{3FHL~J0137.9+5815}: %
Our spectrum (S/N$\sim$145) is dominated by non-thermal continuum.
We detect very weak absorption features from the host galaxy (Ca~II~3934,3968, Mg~I~5175 and Ca+Fe~5269), yielding a z~=~0.2745, and from the spectral decomposition we obtain N/H$\sim$3.
No previous spectrum is available in literature.

\item[] \textbf{3FGL~J0141.4-0929}: %
The only feature in our spectrum is an absorption doublet at $\sim$4200~$\textrm{\AA}$, which is due to intervening Mg~II~2800, yielding a spectroscopic lower limit of the redshift of z~$\geqslant$~0.501.
This feature is also present in the spectra reported in \citet{stickel1993} and \citet{stocke1997}. 
Note that we do not confirm the emission of Mg~II~2800 and [O~II]~3727 at z~=~0.737 proposed by \citet{stocke1997}.

\item[] \textbf{3FHL~J0148.2+5201}: %
In our spectrum, the Ca~II~3934,3968 doublet is apparent at $\sim$5670~$\textrm{\AA}$ and, together with the G-band~4305 absorption line at 6183~$\textrm{\AA}$, yield z~=~0.437. 
The decomposition of the spectrum provides a N/H$\sim$5, and $\alpha$=1.3 for the nuclear power law component.
The spectrum published by \citet{alvarez2016a}, with a poorer S/N, is featureless.

\item[] \textbf{3FHL~J0241.3+6543}: %
We detect two emission lines at $\sim$7370~$\textrm{\AA}$ which can be interpreted as H$_{\alpha}$~6563 and [N~II]~6584 at z~=~0.1211 (see the close-up in Fig.~\ref{fig:closeup}). 
At the same redshift, there is a further emission doublet that we attribute to [S~II]~6716,6731, and an absorption line due to Na~I~5892.
We note that \citet{marchesi2018}, on the basis of an absorption doublet at $\sim$4600~$\textrm{\AA}$, identified as Mg~II, suggest a redshift lower limit z$\geq$0.645. This conflicts with our redshift determination.

\item[] \textbf{3FHL~J0250.5+1712}: %
We obtain a spectrum with S/N$\sim$100 which exhibits clear absorption lines attributed to the host galaxy (Ca~II~3934,3968, H$_{\delta}$~4861, G-band~4305, Mg~I~5175, H$_{\beta}$~4861 and Na~I~5892). 
This yields a redshift of z~=~0.2435, confirming the proposal of \citet{archambault2016}.
For the emission lines, the upper limit of EW is $\sim$~0.45~$\textrm{\AA}$.
The present of the host galaxy is well detected in the decomposition reported in Fig.~\ref{fig:decomposition}, where a N/H~=~2 is estimated.

\item[] \textbf{3FHL~J0322.0+2336}: %
Our optical spectrum, with S/N$>$160, is largely superior to one reported by \citet{lau1998}, but still no significant features are present.
The upper limit for the emission/absorption lines is EW~=~0.10~-~0.25~$\textrm{\AA}$, corresponding to a lower limit of the redshift of z~$>$~0.25.

\item[] \textbf{3FHL~J0423.8+4149}: %
We detect a single emission line at 6997.6~$\textrm{\AA}$ of equivalent width EW~=~1.9~$\textrm{\AA}$. 
The most plausible identification of this line is [O~III]~5007 that yields a tentative redshift of z~=~0.3977. 
The other component of [O~III] 4959 doublet is washed out by the telluric band at 6870~$\textrm{\AA}$. 
This target was proposed as neutrino source by \citet{righi2019mwl} and \citet{paiano2019ATel4c41}.

\item[] \textbf{3FHL~J0433.1+3227}: %
Even though our optical spectrum (S/N$\sim$100) is of higher quality than that presented by \citet{alvarez2016a} (S/N$\sim$5), we still do not find significant features and from the minimum detectable EW~$\sim$~0.3~$\textrm{\AA}$, we can set a redshift lower limit z~$>$~0.45.

\item[] \textbf{3FHL~J0433.6+2905}: %
The spectrum is severely reddened with \textit{E(B-V)}~=~0.66. 
There is a possible broad emission feature centered at $\sim$5340~$\textrm{\AA}$ in a region of the spectrum where the S/N~$\sim$15. 
A similar feature maybe is present in the spectrum reported in \citet{shaw2013}, but not identified by the authors. 
Supposing that this feature is due Mg~II~2800, the redshift would be $\sim$0.91.

\item[] \textbf{3FHL~J0434.7+0921}: %
In our spectrum no spectral feature is apparent.
We set a lower limit of z~$>$~0.1 from the lack of absorption lines from the host galaxy. 

\item[] \textbf{3FHL~J0500.3+5238}: %
The target is highly reddened with \textit{E(B-V)=0.75}. 
Several photospheric absorption lines are apparent (G-band~4305, H$_{\beta}$~4861, Mg~I~5175, Na~I~5892 and H$_{\alpha}$~6564), yielding z~=~0.1229. 
At the same redshift there is a narrow emission line attributed to [N~II]~6584 with EW$\sim$1.3~$\textrm{\AA}$.
The spectral decomposition in the power law ($\alpha\sim$-1.0), due to the nucleus, and in the host galaxy template is reported in Fig.~\ref{fig:decomposition} and we find a N/H~=~0.1.

\item[] \textbf{3FHL~J0506.0+6113}: %
The optical spectrum (S/N$\sim$70) is highly reddened with \textit{E(B-V)}~=~0.55. 
We find a hint of Ca~II~3934,3968 doublet at 6042,6096~$\textrm{\AA}$ that would correspond to z$\sim$0.54.

\item[] \textbf{3FHL~J0515.8+1528}: %
No significant feature is apparent in our spectrum (S/N$\sim$130). 
From the null detection of absorption lines of the host galaxy, we can set a lower limit of z~$>$~0.2.

\item[] \textbf{3FHL~J0540.5+5823}: %
There are no convincing emission/absorption features in our spectrum (S/N$\sim$60). 
A lower limit of z~$>$~0.10 can be set from the lack of detected absorption lines of the host galaxy.
This source is inserted in the \citet{padovani2016} list of hard gamma-ray sources found around the position of IceCube events.

\item[] \textbf{3FHL~J0600.3+1245}: %
This gamma-ray object, associated to the radio source NVSS~J060015+124344, is classified as BCU in the \textit{Fermi} catalogs. 
It is spatially coincident with the X-ray source 1RXS~J006014.8+124341 and with the optical counterpart at (RA,DEC) = (06:00:15, 12:43:43).
Our optical spectrum exhibits several absorption lines (G-band~4305, H$_{\beta}$~4861, Mg~I~5175, Na~I~5892 and H$_{\alpha}$~6563) attributed to the old stellar population of the host galaxy, that allow us to measure z~=~0.0835. 
At the same redshift, we also detect, at $\sim$7133~$\textrm{\AA}$, two narrow emission lines (EW~=~1.2~$\textrm{\AA}$ and EW~=~1.8~$\textrm{\AA}$) due to H$_{\alpha}$~6563 and [N~II]~6584 (see the close-up in Fig.~\ref{fig:closeup}).

\item[] \textbf{3FHL~J0601.0+3837}: %
The source was proposed as a BLL by \citet{paggi2014} who show a featureless spectrum of the X-ray counterpart.
The source is faint (g~=~20.5) and severely extincted (\textit{E(B-V)}~=~0.46), and we obtained a spectrum with a S/N$\sim$40.
This is basically featureless, apart for a possible doublet (at $\sim$6560~$\textrm{\AA}$) that, if attributed to Ca~II doublet of the host galaxy, corresponds to a redshift z~=~0.662.

\item[] \textbf{3FHL~J0602.0+5316}: %
The gamma-ray source is associated to the bright radio galaxy GB6J601+5315 and it is inside the error box of a neutrino event detected by IceCube \citep{padovani2016}. 
No previous optical spectrum is found in literature. 
Our spectrum of the source (g~=~17.0) shows the absorption lines (Ca~II, G-band, etc...) of the galaxy stellar population at z~=~0.0522. 
The decomposition of the optical spectrum (see Fig.~\ref{fig:decomposition}) in an elliptical template and power-law ($\alpha\sim$ -1.0) emission indicates the presence of a \textbf{non-thermal} component contributing for $\sim$45\%. 

\item[] \textbf{3FHL~J0607.4+4739}: %
An optical spectrum, obtained by \citet{shaw2013}, does not show any significant features. 
From our lineless spectrum (S/N~=~100), on the basis of the lack of detectable galaxy absorption lines, we can set a redshift lower limit z~$>$~0.10.

\item[] \textbf{3FHL~J0612.8+4122}: %
Our spectrum is dominated by a power law continuum without significant intrinsic emission/absorption lines. 
However, we clearly detect intervening absorption systems of Mg~II~2800, Fe~II~2382, and Fe~II~2586,2600 at z~=~1.107. 
The Mg~II absorption line doublet is also apparent in \citet{shaw2013}.

\item[] \textbf{3FHL~J0620.6+2645}: %
This gamma-ray source (g~=~18.5) is classified as BCU in the \textit{Fermi} catalogs. 
No previous optical spectrum is available.   
In our spectrum, many absorption lines from the host galaxy are clearly detected, yielding z~=~0.1329.
The decomposition of the spectrum, in a elliptical galaxy and a power law ($\alpha$~=~-1.5) emission, allows us to set N/H$\sim$0.5.

\item[] \textbf{3FHL~J0640.0-1254}: %
No previous redshift is reported in literature. 
Our spectrum exhibits absorption lines (Ca~II, G-band, Mg~I, Na~I) of the host galaxy diluted by the non thermal emission from the nucleus, allowing to measure z~=~0.1365. The spectral decomposition in the elliptical galaxy and nucleus power law component provides an N/H$\sim$ 1 (see Fig.~\ref{fig:decomposition}).
In addition to the absorption feature, we also detect a narrow emission line at 7482~$\textrm{\AA}$ (EW~=~0.6~$\textrm{\AA}$) corresponding to [N~II]~6584.
We note that the object visible at 6.5" from the target is a star and it is derived by the GTC spectrum obtained during the same observation of our target.

\item[] \textbf{3FHL~J0702.6-1950}: %
No optical spectra are found in literature. 
We obtain the spectrum orientating the slit to intersect the target (g~=~19.1) and the close object at $\sim$5", which is a star (the classification is provided by GTC spectroscopy). 
The moderate S/N$\sim$50 spectrum of the $\gamma$-ray target is dominated by a featureless continuum characterized by a typical power law shape with spectral index $\alpha$~=~-1.4.
Based on the lack of spectral features from the host galaxy, we set a lower limit of z~$>$~0.10.

\item[] \textbf{3FHL~J0706.5+3744}: %
We detect a faint absorption doublet at $\sim$4350~$\textrm{\AA}$ (EW$\sim$0.6~$\textrm{\AA}$) that is attributed to Ca~II 3934,3968 (see Fig.~\ref{fig:closeup}) yielding the redshift z~=~0.1042. 
If these lines were ascribed to the starlight of the host elliptical galaxy, we would expect to observe the trace of the host galaxy in the continuum, which is not present. 
Moreover these detected lines are narrower, compared to the typical Ca~II line width from elliptical galaxies, rather indicating an interstellar absorption origin. 
Indeed at $\sim$12" from the target, there is another galaxy for which we obtained an optical spectrum: we found several emission lines due to H$_{\beta}$~4861, [OIII]~4959,5007 and H$_{\alpha}$~6563 at z~=~0.1042, the same value of the BLL. 
We suppose that the halo gas of this close galaxy (located at the projected distance $\sim$23~kpc at that redshift) can be responsible of the absorption doublet found in the spectrum of $\gamma$-ray source. 
Therefore, for 3FHL~J0706.5+3744, we set the spectroscopic lower limit of the redshift z~$\geq$~0.1042.  

\item[] \textbf{3FHL~J0708.9+2240}: %
This object is a BCU gamma-ray emitter associated to the radio source GB6J0708+2241. 
The optical counterpart (g~=~17.4) is at (RA,DEC)=(07:08:58.3, 22:41:36.0). 
\citet{massaro2015} report an optical spectrum without evident features.
We detect absorption lines (Ca~II, G-band, and Mg~I) due to the old stellar population of the host galaxy, allowing us to locate the source at z~=~0.2966. 
The spectral decomposition, reported in Fig.~\ref{fig:decomposition}),  indicates the presence of a no-thermal power law component ($\alpha$=-1.15) and a N/H~=~5.5. 

\item[] \textbf{3FHL~J0709.1-1525}: %
In the \textit{Fermi} catalog, this BCU $\gamma$-ray source is associated to the radio source PKS~0706-15. 
From the analysis of \textit{Swift}/XRT data, inside the Fermi error box we detect the X-ray source J070912-152703, spatially coincident with the radio counterpart and the optical counterpart (g~=~18.9) at (RA,DEC)=(07:09:12.3,-15:27:00.0). This confirms the association with PKS~0706-15.
No previous spectrum is reported in literature.
Our spectrum, highly reddened (\textit{E(B-V)}~=~0.55), exhibits the shape of an elliptical galaxy with evident absorption lines (Ca~II, G-band, H$_{\beta}$, Ca+Fe, Na~I, and H$_{\alpha}$) at redshift z~=~0.1420. 
In addition we detect an emission line (EW~=~1.0~$\textrm{\AA}$) at 7518~$\textrm{\AA}$ which is attributed to [N~II]~6584~$\textrm{\AA}$.

\item[] \textbf{3FHL~J0723.0-0732}: %
The blazar classification of this $\gamma$-ray emitter is proposed by \citet{marti2004} who report a featureless optical spectrum.
In our spectrum (S/N$\sim$170), we detect the absorption doublet attributed to Ca~II~3934,3968 and a faint feature due to G-band. 
This yields a redshift of z~=~0.3285. 
The decomposition of the spectrum (see Fig.~\ref{fig:decomposition}) in an elliptical template and power-law emission indicates the presence of no-thermal component and allows us to set N/H$\sim$5.

\item[] \textbf{3FHL~J0811.9+0237}: 
There are not previous spectra available for this object.
Our spectrum clearly exhibits absorption lines (Ca~II, G-band, H$_{\beta}$, Mg~I, Ca+Fe, and Na~I ) due to the host elliptical galaxy at z~=~0.1726. 
In addition a faint emission line attributed to [N~II]~6584 is detected at the same redshift. 
The decomposition of the spectrum in an elliptical galaxy and power-law ($\alpha$=-1) emission reveals the presence of no-thermal component and the N/H$\sim$0.75 (see Fig.~\ref{fig:decomposition}).
We note that 3FHL~J0811.9+0237 could belong to a small group of galaxies. 
There are two neighbour galaxies at $\sim$9" (on East) and $\sim$22" (North-West) at redshift similar to our target (see Fig.~\ref{fig:0811}): from our GTC spectroscopy of G1 and from the SDSS spectrum available for G2, we find that these companions are elliptical galaxies at z~=~0.1697 of projected velocity of $\sim$900~km~s$^{-1}$ relative to our target.

\setcounter{figure}{4}
\begin{figure}
\centering
\includegraphics[width=0.35\textwidth]{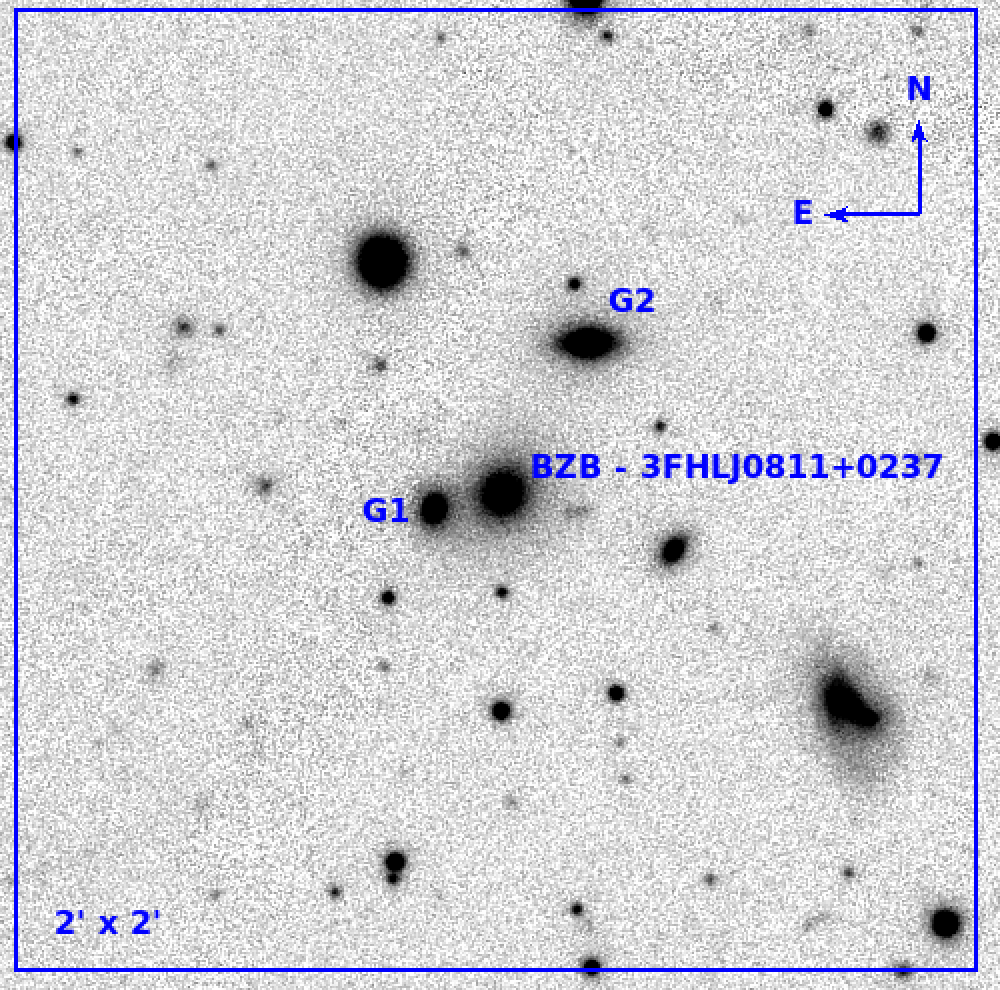}
\includegraphics[width=0.25\textwidth, angle=-90]{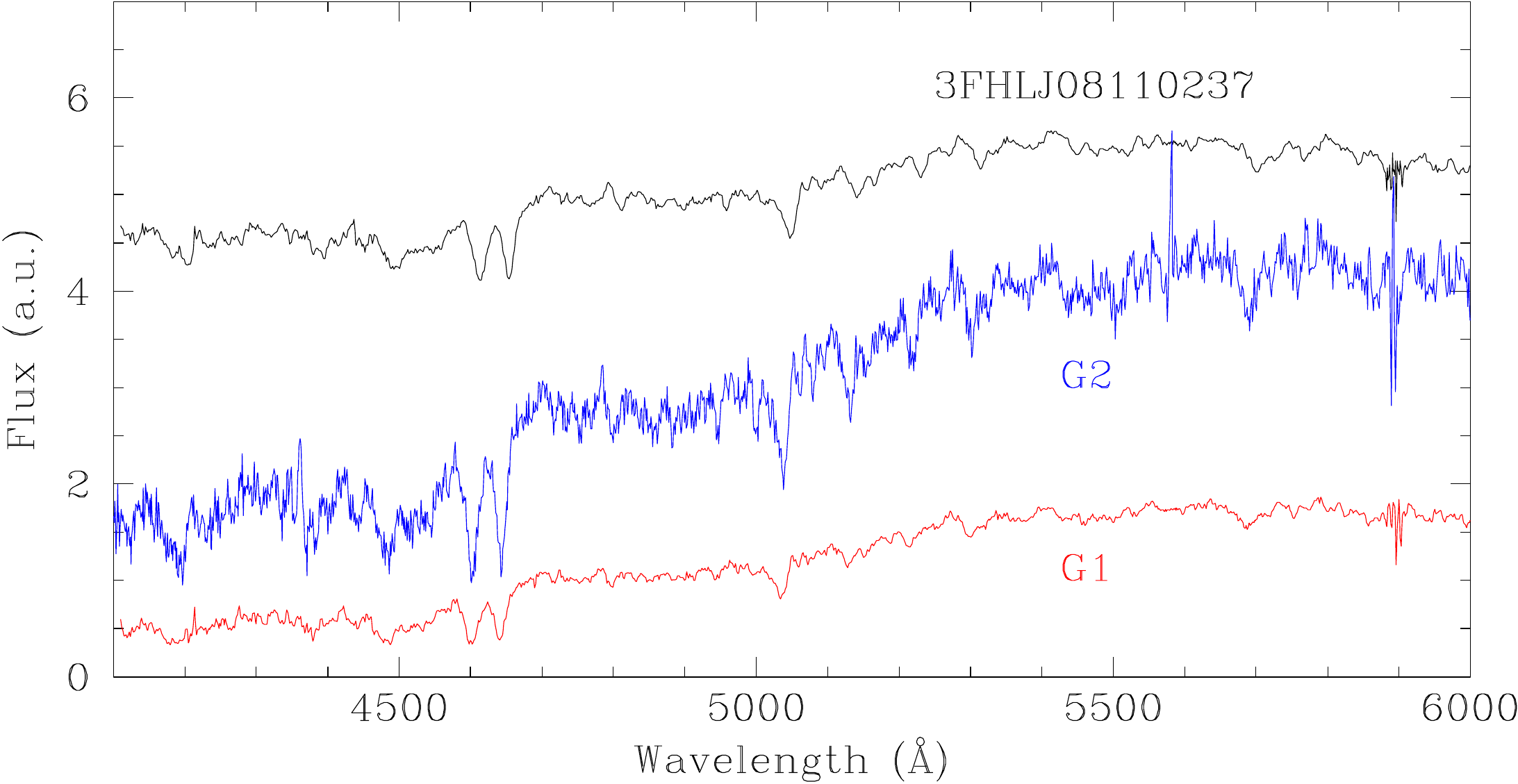}
\caption{\textit{Upper panel}: The PANSTARR \textit{i}-band image (North up and East left) of the BLL 3FHL~J0811.9+0237 (in the center). There are two close companions of the BLL: G1 at $\sim$9" and G2 at $\sim$22". \textit{Bottom panel}: Comparison of the BLL spectrum with the spectra of the two companions (see details in the text).}
\label{fig:0811}
\end{figure}

\item[] \textbf{3FHL~J0816.4-1311}: %
Despite the high quality optical spectrum (S/N$\sim$300), no emission/absorption lines are found. 
On the basis of the minimum detectable EW ($\sim$0.10~$\textrm{\AA}$) and the lack of feature detection from the host galaxy, we can set a redshift lower limit of z~$>$~0.40.
There are previous spectra reported in literature by \citet{shaw2013}, who show a featureless spectrum, and by \citet{pita2014} that found intervening absorption systems attributed to Mg~II~2800 (at z~=~0.1902, z~=~0.2336, and z~=~0.2882) in a spectral range not covered by our spectrum.

\item[] \textbf{3FHL~J0905.5+1357}: %
We observed this source at two different epochs separated by an interval of about one year (see Table~\ref{tab:sample}). 
It is noticeable that the source magnitude obtained from the acquisition images varied by $\sim$~one mag. 
In Fig.~\ref{fig:spectra} we report the average spectrum of the two epochs. 
The shape is the typical BLL continuum and the only intrinsic feature detected is an emission line (average EW$\sim$0.7~$\textrm{\AA}$) at 6133~$\textrm{\AA}$ (see the close-up around this line in Fig.~\ref{fig:closeup}) which, if attributed to [OIII]~5007, yields a redshift of 0.2239. 
The line luminosity is $\sim$8$\times$10$^{40}$~erg/s during the two states, while the continuous varies of a factor of $\sim$3.
This source is quoted in \citet{padovani2016} as spatially coincident with the error box of a neutrino IceCube event and there are two previous spectra by \citet{shaw2013} and the SDSS. 
Both spectra appear featureless and correspond to a high state of the source (F(6100$\textrm{\AA}$)$\sim$1$\times$10$^{-16}$~erg cm$^{-2}$ s$^{-1}$) that probably has hampered the detection of the line, which appeared clearly in our observations when the source was in a lower flux state.  

\item[] \textbf{3FHL~J0910.5+3329}: %
Our optical spectrum with S/N$\sim$200 appears featureless, in agreement with the previous spectra available in literature \citep[][and that provided by the SDSS survey]{bauer2000, shaw2013}. 
We can set a redshift lower limit z~$>$~0.15 on the basis of the non-detection of features due to the host galaxy.

\item[] \textbf{3FHL~J0953.0-0840}: %
The only spectrum available in literature is provided by \citet{shaw2013} that is featureless.
In our good optical spectrum (S/N$\sim$220), still no emission/absorption line is detected. 
The spectrum exhibits the power-law shape, typical for BLLs, with spectral index of $\alpha$~=~-1.5. 
By the minimum EW method, we can set the redshift lower limit z~$>$~0.15.

\item[] \textbf{3FHL~J1037.6+5711}: %
All previous spectra reported in literature \citep[][and the SDSS survey]{lau1998, caccianiga2002, shaw2013} do not exhibit emission/absorption lines.  
Our high quality (S/N$\sim$300) spectrum shows the typical BLL power-law shape and it still is featureless. 
The estimated minimum EW is $\sim$0.15~$\textrm{\AA}$ and this yields a redshift lower limit z~$>$~0.25. 

\item[] \textbf{3FHL~J1055.6-0125}: %
No redshift is reported in literature.
The spectral shape of our spectrum is well described by a power law emission ($\alpha\sim$~-1.5), typical of a BLL. 
No emission or absorption features are apparent at level of EW$\sim$0.40~$\textrm{\AA}$ and we can set a redshift lower limit of z~$>$~0.55.

\item[] \textbf{3FHL~JJ1059.1-1134}: %
The source was observed by \citet{landoni2013} and \citet{shaw2013} and they did not detect any features. 
Also our spectrum appears featureless.
The minimum detectable EW spans between 0.30~$\textrm{\AA}$ and 1.15~$\textrm{\AA}$ along the spectrum, allowing us to set a redshift lower limit based on the lack of galaxy absorption lines of z$>$0.10.

\item[] \textbf{3FHL~J1150.5+4154}: %
There is an SDSS spectrum of the source and one in the collection of \citet{shaw2013}. 
Neither absorption nor emission features are apparent.
Also our high S/N$\sim$200 spectrum is featureless and from the minimum EW$\sim$0.25~$\textrm{\AA}$, we can set a lower limit z~$>$~0.25 from the lack of detection of features from the host galaxy.

\item[] \textbf{3FHL~J1233.7-0145}: %
Optical spectra provided by \citet{shaw2013}, \citet{kugler2014} and from the SDSS survey are featureless and no spectroscopic redshift is provided.
Also in our spectrum no emission/absorption features are detected and, from a minimum detectable EW$\sim$0.45~-~0.95~$\textrm{\AA}$, we can set a redshift lower limit of 0.10. 

\item[] \textbf{3FHL~J1253.1+5300}: %
In our high quality spectrum (S/N$\sim$250) of the target (g~=~16.6), we detect a faint absorption doublet system (EW$\sim$0.3~$\textrm{\AA}$) at $\sim$4660~$\textrm{\AA}$ attributed to intervening Mg~II~$\textrm{\AA}$ cold gas, allowing us to set the spectroscopic redshift lower limit z~$\geq$~0.6638. 
The same line system is found in the spectrum of the SDSS survey and reported by \citet{shaw2013}. 

\item[] \textbf{3FHL~J1418.4-0233}: %
All optical spectra reported in literature \citep[][and that provided by SDSS survey]{shaw2013,kugler2014} appear featureless and no redshift is measurable.
Our S/N$\sim$200 optical spectrum of this bright target (g~=~16.4) does not exhibit emission or absorption lines and the minimum detectable EW is $\sim$0.15-0.30~$\textrm{\AA}$. 
We can set a redshift lower limit of z~$>$~0.12.

\item[] \textbf{3FHL~J1445.0-0326}: %
This object has been classified as BLL by \citet{bauer2000}, but its redshift was not determined because of the absence of emission and absorption features. 
Featureless spectra are reported in \citet{sbarufatti2006a} and \citet{piranomonte2007}
Also our spectrum (S/N~=~200) does not show apparent spectral lines at level of EW$\sim$0.20~$\textrm{\AA}$ and we can set a redshift lower limit of z~$>$~0.45.

\item[] \textbf{3FHL~J1447.9+3608}: %
In our spectrum (SNR$\sim$250), there is a clear absorption doublet at 4865~$\textrm{\AA}$, which is due to intervening Mg~II~2800 system, yielding a spectroscopic lower limit of z~$\geq$~0.738. 
The feature was noted by \citet{shaw2013} and appears also in the SDSS spectrum. 

\item[] \textbf{3FHL~J1454.5+5124}: %
The redshift z~=~1.0831 reported in literature was proposed by the automatic procedure of the SDSS survey. 
However no convincing line identification is provided and also the spectrum shown by \citet{shaw2013} appears featureless.
In our spectrum of very good quality, with S/N$\sim$320, no spectral features are found. 
We can set a redshift lower limit of z~$>$~0.40 by the minimum EW$\sim$0.10~$\textrm{\AA}$. 

\item[] \textbf{3FHL~J1503.7-1541}: %
The optical spectra reported in literature \citep{bauer2000, sbarufatti2006a, shaw2013} are featureless.
Also our spectrum does not present emission/absorption lines and we can set a redshift lower limit of z~$>$~0.10.

\item[] \textbf{3FHL~J1549.9-0659}: %
This gamma-ray emitter is associated to the radio source NVSS~J154952-065907 and proposed as blazar candidate in the \textit{Fermi} catalog because no spectra are available in literature.
Our spectrum establishes the BLL nature of the source and we find faint absorption lines due to Ca~II~3934,3968 and G-band~4305, attributed to the stellar population of the host galaxy, at a z~=~0.418.

\item[] \textbf{3FHL~J1748.6+7005}: %
In literature, \citet{stickel1989} provided a tentative redshift of z~=~0.77 based on identification of the very faint (EW~=~0.4~$\textrm{\AA}$) [O~II] emission line at 6600~$\textrm{\AA}$. 
This value was subsequently confirmed by \citet{lawrence1996} from the detection of additional lines due to C~III] (EW~=~0.4~$\textrm{\AA}$) and [O~III] (EW~=~0.7~$\textrm{\AA}$).
Our high S/N ($\sim$300) spectrum of the bright target (g~=~16.6) is featureless. 
Note that we cannot detect the emission lines claimed in the previous works because they are out of our spectral range.

\item[] \textbf{3FHL~J1800.5+7827}: %
We clearly detect a significant (EW~=~8.3~$\textrm{\AA}$) emission line at 4712~$\textrm{\AA}$ (see the close-up in Fig.~\ref{fig:closeup}) that, if attributed to Mg~II~2800, leads to a redshift z~=~0.683. 
From the literature, a similar value is reported in \citet{hewitt1993} and in \citet{stickel1994} where neither spectrum nor line identification are shown. 

\item[] \textbf{3FHL~J1841.3+2909}: %
This 3FHL source is associated to the radio source MG3~J184126+2910 and proposed as BLL candidate \citep{massaro2013}. 
Our de-reddened (\textit{E(B-V)}=0.21) optical spectrum is described as a power-law ($\alpha$~=~-1.1) and no evident lines are found. 
We set a redshift lower limit of z$>$~0.1.
There is also another spectrum published by \citet{marchesi2018} that is again featureless.

\item[] \textbf{3FHL~J1904.1+3627}: %
This target is a candidate blazar in the \textit{Fermi} catalog and associated to the radio source MG2~J190411+3627. 
In \citet{marchesi2018} the optical spectrum seems featureless, mainly due to a poor S/N.
Our spectrum (S/N$\sim$60) is modelled by an elliptical galaxy shape and shows several absorption lines due to the old stellar population of the host galaxy. The redshift of the target is z~=~0.08977.
From the spectral decomposition (see Fig.~\ref{fig:decomposition}) we can determine that the AGN contributes to $\sim$15\% (N/H=0.2).

\item[] \textbf{3FHL~J1911.5-1908}: %
This source is a BCU of the \textit{Fermi} catalog associated to the radio source PMN~J1911-1908. 
\citet{marchesi2018} report an optical spectrum of modest S/N that appears featureless.
In our spectrum, several absorption lines are found due to the old stellar population (Ca~II, G-band, Mg~I, Ca+Fe and Na~I), yielding a redshift z~=~0.138.  
At the same redshift, we also detect narrow emission lines due to forbidden transitions attributed to [O~II]~3727, [O~III]~5007, [N~II]~6548, H$_{\alpha}$~6563 partially blended, and [N~II]~6584. 

\item[] \textbf{3FHL~J1921.8-1607}: %
Our good (S/N$\sim$125) optical spectrum does not exhibit emission or absorption lines. From the lack of detection of features of the host galaxy, with the minimum detectable EW$\sim$0.20-0.50~$\textrm{\AA}$ and magnitude g~=~17.6, we can set a lower limit of the redshift z~$>$~0.12.
In literature, there is another spectrum in \citet{shaw2013}  that is featureless.
\end{itemize}


\section{Summary and Conclusions} 
\label{sec:notes}
From our optical  spectroscopy, we can assess that all sources in the observed data set are BLL. 
We determined the redshift for 26 objects from absorption lines of their host galaxies and in some cases also from emission lines. 
For other 5 objects we found a spectroscopic lower limit from the detection of intervening absorptions. The remaining 24 sources have featureless optical spectra thus we estimate a redshift lower limit   from the absence of absorption features of the host galaxy.  
Moreover for all the latter sources there are not intervening absorptions that suggests they are at relatively low redshift. Considering the absorptions from Mg~II we expect that the number of Mg~II absorbers per redshift bin (dN/dz), with EW$>$0.3~$\textrm{\AA}$, is $\sim$0.5 \citep{zhu2013}. Since none are detected above 4100~$\textrm{\AA}$, one must have on average $\bar z$~<~0.6.
This is consistent with the median value of the detected redshifts ($\bar z\sim$0.3.). 

BLL are characterized by the lack (or very weak) features in their optical spectra. Apart from the absorption lines from the host galaxies, that are always present if sufficient high quality spectrum is available \citep{landoni2014}, it is of interest to evaluate the properties of the possible emission lines compared with other active nuclei.
Weak narrow emission lines are observed in a number of cases for BLL. These can arise either from a region of significant star formation or from nuclear emission in the narrow line region \citep{bressan2006, paiano2017tev, paiano2018txs}. 
The detection of weak broad emission lines is rarer \citep[see e.g.][]{2006bsbarufatti, landoni2012}. It adds an important ingredient for the understanding of this class of active galatic nuclei \citep[e.g.][]{giommi2013}, in particular in the connection with the FSRQ and the possibility of an evolutionary relation between these two classes \citep[for example][]{giommi2013, ajello2014}.

In 4 objects we detected weak (EW$\sim$0.5-1.9~$\textrm{\AA}$) [O~III]~5007 emission lines (see Table~\ref{tab:ew}). This yields an average line luminosity \textit{L}~$\approx$~8~$\times$10$^{40}$~erg/s. 
For the 13 objects of known redshift and for which [O~III] would be observable in our spectra, on average we set an upper limit  line luminosity of 3$\times$10$^{40}$ erg/s. 
A similar limit was also derived for the 24 sources of unknown redshift, assuming that the [O~III] emission line be inside the observed range. 
In addition to the [O~III] emission line, we also detect [N~II]~6584 emission feature in 7 targets (only in one of these cases also [O~III] is present). The derived [N~II] line luminosity is in the range 0.1-2$\times$10$^{41}$ erg/s. 
Only for two sources we detected measurable H$_{\alpha}$ emission. These weak (EW$\sim$1.5~$\textrm{\AA}$) lines imply H$_{\alpha}$ luminosity of 2.5$\times$10$^{40}$ and 1.2$\times$10$^{41}$~erg/s.  
At face value, the [O~III] line luminosity is significantly lower than that found for the low redshift (0.1$<$$z$$<$0.5) QSOs \citep[see e.g.][]{shen2011} at comparable observed continuum luminosity (see Fig.~\ref{fig:plotlines}-\textit{upper panel}-a). 
On average the [O~III] luminosity for our BLL sample is a factor $\sim$50 lower than that found for the QSOs ($\sim$10$^{42}$ erg/s) and a factor $\sim$20 lower considering the [O~III] luminosity distribution of the type~II QSOs \citep[see ][]{zakamska2003}.
This might indicate either that the lines are indeed fainter of that found in the QSOs at the same continuum level or that the continuum is significantly enhanced by the presence of relativistic beaming that is characteristic of this kind of sources \citep[e.g.][]{urry1995}. 
Actually, assuming a beaming factor of $\sim$10, both measured values and upper limits move to the region at the lowest line and continuum luminosities as it is displayed in Fig.~\ref{fig:plotlines}-\textit{upper panel}-b). 

Another important result of our spectroscopy is that we do not find broad emission lines, except for 3FHL~J1800.5+7827 for which we detect Mg~II emission line of EW~=~8.3$\textrm{\AA}$, FWHM$\sim$3500km/s and L$_{MgII}$~=~2.2$\times$10$^{43}$ erg/s. 
This is at the limit of the distribution of the QSOs at similar redshift and luminosity of the continuum (see Fig.~\ref{fig:plotlines}-\textit{bottom panel}-c). 
Among our BLL with known redshift, only two are at z$>$0.5 and the Mg~II emission could occur in the observed spectral range. For these sources we estimate an upper limit of the Mg~II luminosity $\leq$~2$\times$10$^{42}$ erg/s, that is a factor 20 smaller than the average value of the QSOs reported in \citet{shen2011}. 
A similar value is also found for the BLL for which we provide a spectroscopic redshift lower limit.  
As for the case of [O~III] lines, when accounting for the same beaming correction data points and limits locate in the region of the lowest luminosity (see Fig.~\ref{fig:plotlines}-\textit{bottom panel}-d).

The comparison and contrast of the properties of the emission lines of BLL and AGN obviously requires a thorough and extended discussion, which deserve the comparison of large and homogeneous samples to properly investigate this issue.

\begin{figure*}
\centering
\includegraphics[width=0.8\textwidth]{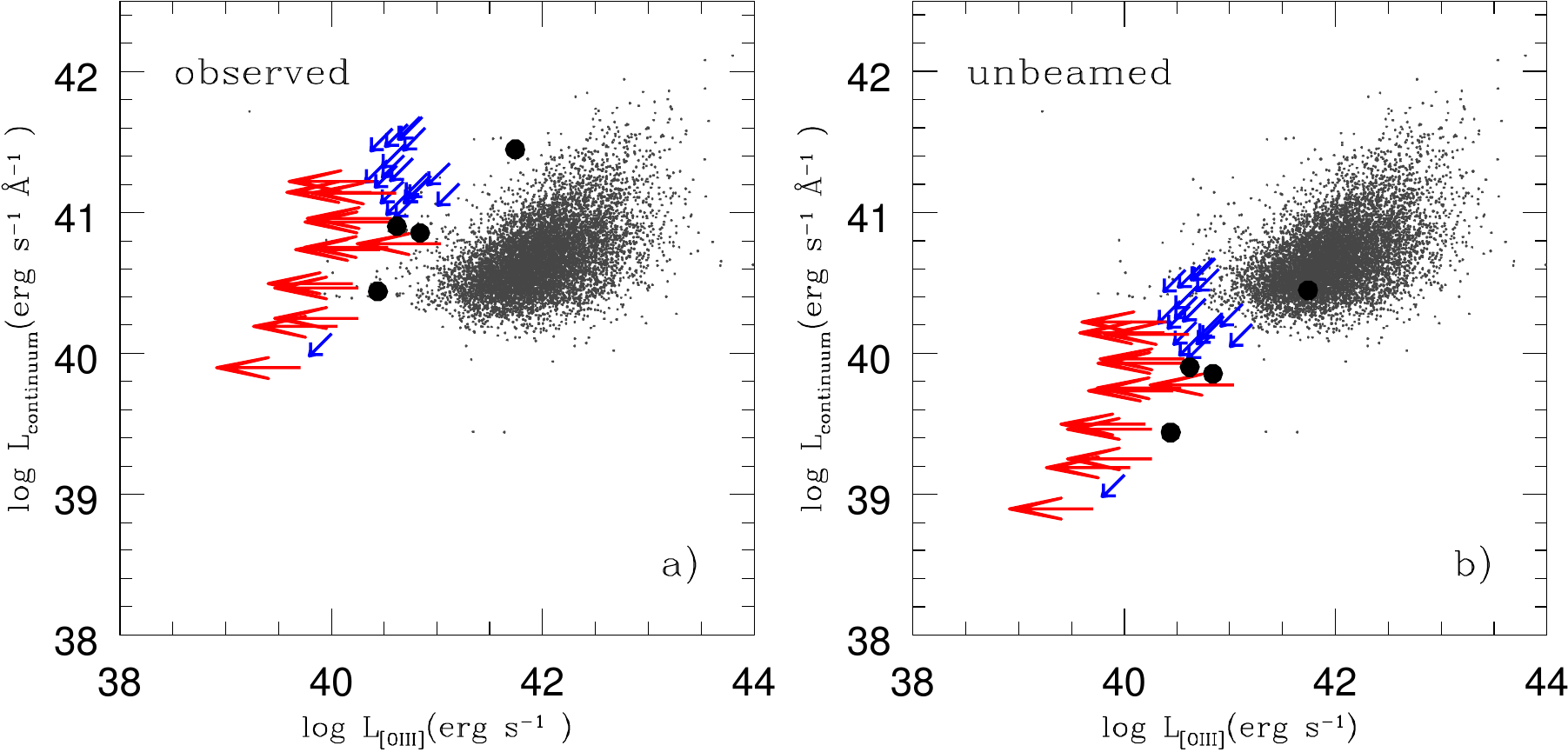}
\includegraphics[width=0.8\textwidth]{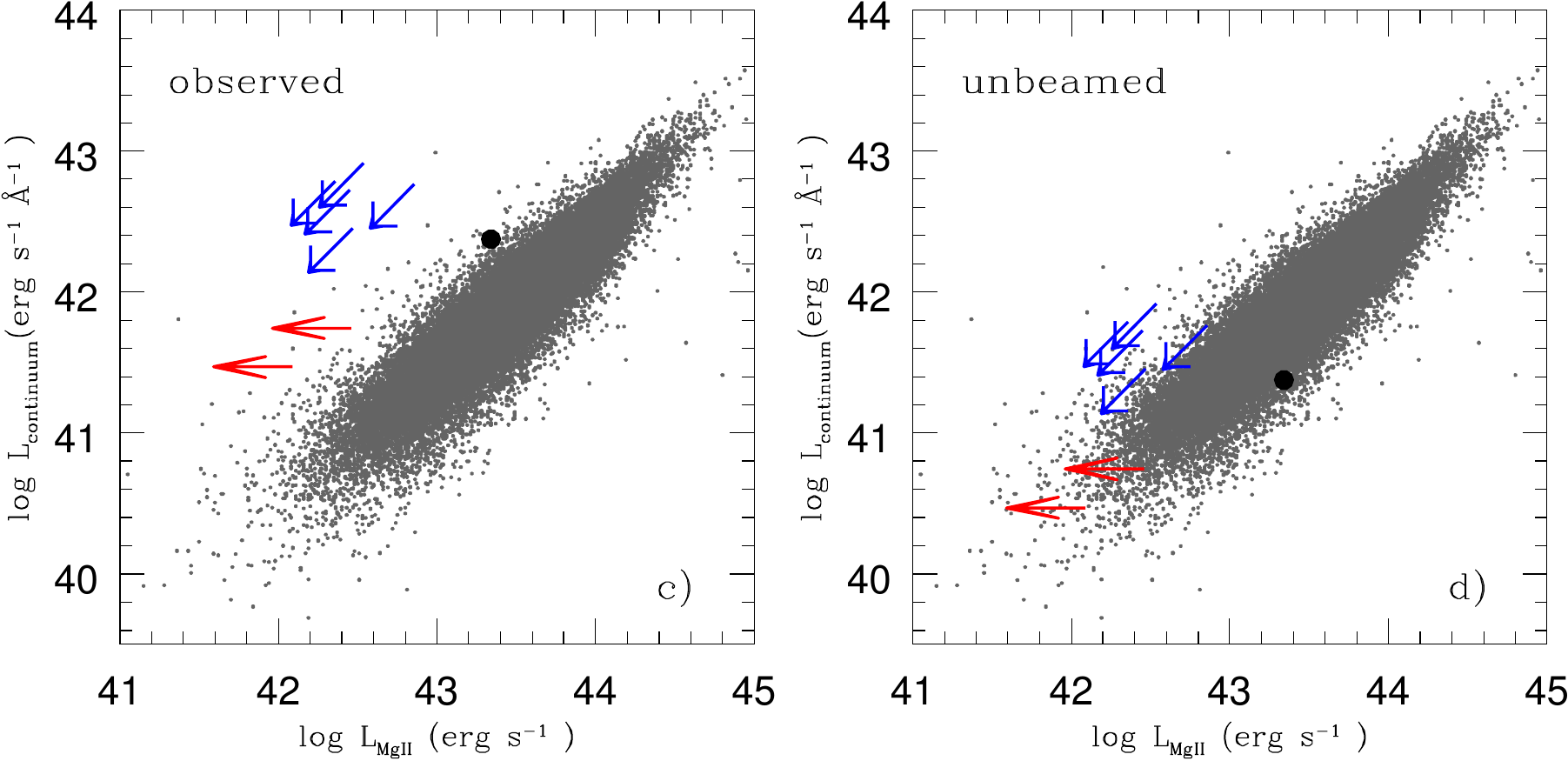}
\caption{\textit{Upper panels}: The relation between the [O~III] line luminosity and the beamed (on the left) and unbeamed of a factor ten (on the right) continuum luminosity of our sample of 3FHL BLL compared to that of low redshift (0.1$<$$z$$<$0.5) QSOs from SDSS (grey points) taken from \citet{shen2011}. The four cases with detected [O~III] emission line are marked with black circles (see also Table~\ref{tab:ew}). The 13 BLL with known redshift but without the [O~III] line detection are indicated as red arrows and represent upper limits for L$_{[OIII]}$. Finally the blue arrows are the upper limits for [O~III] and continuum luminosity for the BLL without redshift (see text for details). \textit{Bottom panels}: The relation between the Mg~II line luminosity and the beamed (on the left) and unbeamed of a factor ten (on the right) continuum luminosity of our BLL sample compared to that of QSO (grey points) with redshift 0.46$<$$z$$<$1.5 from \citet{shen2011}. The only case with Mg~II line detected is marked with black circle, the two sources with z$>$0.5 and without Mg~II line detection are marked as red arrows, while the BLL with spectroscopic redshift lower limit are labelled with blue arrows.  
}

\label{fig:plotlines}
\end{figure*}

\setcounter{table}{0}
\begin{table*}
\begin{center}
\caption{THE SAMPLE OF THE 3FHL TEV CANDIDATES OBJECTS AND JOURNAL OF THE GTC OBSERVATIONS }\label{tab:sample}
\begin{tabular}{llllllllll}
\hline 
3FGL Name  &    Counterpart     &  RA              &    DEC             & Class &  E(B-V)  &  Obs. date  & t$_{Exp}$      &   Seeing       & g \\   
           &                            &  (J2000)         &        (J2000)     &       &          &       &  (s)      &   ('')         &  \\ 
\hline
3FHLJ0009.4+5030 & NVSSJ000922+503028   & 00:09:22.8 &  50:30:28.8 & bll    & 0.13 & 06-11-2018 & 4500 & 1.5 & 18.9 \\
3FHLJ0015.7+5551 & GB6J0015+5551        & 00:15:40.1 &  55:51:45.0 & bll    & 0.37 & 06-11-2018 & 3000  &  1.5 & 18.7   \\
3FHLJ0045.3+2127 & GB6J0045+2127        & 00:45:19.3 &  21:27:40.0 & bll    & 0.03 & 06-11-2018 & 1200  &  1.7 & 17.7   \\
3FHLJ0045.7+1217 & GB6J0045+1217        & 00:45:43.3 &  12:17:12.0 & bll    & 0.07 & 26-11-2018 & 3000  &  2.0 & 17.6   \\ 
3FHLJ0131.1+6120 & 1RXSJ013106.4+612035 & 01:31:07.3 &  61:20:34.0 & bll    & 0.79 & 03-12-2018 & 3600  &  1.3 & 20.3  \\
3FHLJ0134.4+2638 & 1RXSJ013427.2+263846 & 01:34:28.1 &  26:38:43.0 & bcu    & 0.09 & 03-12-2017 & 1000  &  1.3 & 17.6   \\
3FHLJ0137.9+5815 & 1RXSJ013748.0+581422 & 01:37:50.4 &  58:14:11.0 & bll    & 0.45 & 03-01-2019 & 3600  &  1.2 & 18.9   \\
3FGLJ0141.4-0929 & PKS0139-09           & 01:41:25.8 & -09:28:43.7 & bll    & 0.02 & 27-12-2017 & 1350  &  1.1 & 16.8   \\
3FHLJ0148.2+5201 & GB6J0148+5202        & 01:48:20.2 &  52:02:06.0 & bll    & 0.19 & 04-12-2018 & 2400  &  2.7 & 18.1   \\
3FHLJ0241.3+6543 & TXS0237+655          & 02:41:21.6 &  65:43:11.9 & bcu    & 1.09 & 07-12-2016 & 3600  &  0.9 & 20.9   \\
3FHLJ0250.5+1712 & NVSSJ025037+171209   & 02:50:37.9 &  17:12:09.0 & bll    & 0.12 & 30-10-2018 & 1200  &  1.7 & 17.8   \\
3FHLJ0322.0+2336 & MG3J032201+2336      & 03:22:00.0 &  23:36:11.0 & bll    & 0.17 & 30-10-2018 & 1200  &  1.3 & 17.5   \\
3FHLJ0423.8+4149 & 4C+41.11             & 04:23:56.0 &  41:50:02.7 & bll    & 0.63 & 26-03-2018 & 7200  &  1.9 & 20.3   \\
3FHLJ0433.1+3227 & NVSSJ043307+322840   & 04:33:07.7 &  32:28:40.0 & bll    & 0.38 & 08-11-2018 & 3900  &  0.7  & 19.7  \\
3FHLJ0433.6+2905 & MG2J043337+2905      & 04:33:37.8 &  29:05:55.4 & bll    & 0.66 & 21-02-2017 & 4500  &  1.9  & 21.9  \\ 
3FHLJ0434.7+0921 & TXS0431+092          & 04:34:41.0 &  09:23:49.0 & bcu    & 0.22 & 05-12-2017 & 3900  &  1.6  & 18.1  \\
3FHLJ0500.3+5238 & GB6J0500+5238        & 05:00:21.4 &  52:38:02.0 & bcu    & 0.75 & 26-11-2016 & 3600  &  1.4  & 19.6  \\
3FHLJ0506.0+6113 & NVSSJ050558+611336   & 05:05:58.8 &  61:13:36.0 & bll    & 0.55 & 14-12-2018 & 4800  &  2.0  & 19.6  \\
3FHLJ0515.8+1528 & GB6J0515+1527        & 05:15:47.3 &  15:27:17.0 & bll    & 0.47 & 03-03-2019 & 7200  &  1.7  & 18.9  \\
3FHLJ0540.5+5823 & GB6J0540+5823        & 05:40:30.0 &  58:23:38.0 & bll    & 0.34 & 28-01-2018 & 900   &  1.7  & 18.2  \\
3FHLJ0600.3+1245 & NVSSJ060015+124344   & 06:00:15.0 &  12:43:43.0 & bcu    & 0.41 & 28-11-2016 & 3600  &  0.9  & 18.5  \\
3FHLJ0601.0+3837 & B20557+38            & 06:01:02.8 &  38:38:29.0 & bll    & 0.46 & 09/10-03-2019 & 6000 & 1.6 & 20.5  \\
3FHLJ0602.0+5316 & GB6J0601+5315        & 06:02:00.4 &  53:16:00.0 & bcu   & 0.15 & 03-12-2017 & 900   &  1.2  & 17.0  \\
3FHLJ0607.4+4739 & TXS0603+476          & 06:07:23.2 &  47:39:47.0 & bll    & 0.16 & 20-01-2019 & 1200  &  1.1  & 17.2  \\
3FHLJ0612.8+4122 & B30609+413           & 06:12:51.2 &  41:22:37.0 & bll    & 0.17 & 20-01-2019 & 900   &  1.2  & 18.1  \\
3FHLJ0620.6+2645 & RXJ0620.6+2644       & 06:20:40.0 &  26:43:32.0 & bcu  & 0.34 & 09-04-2019 & 3600  &  1.8  & 18.5  \\
3FHLJ0640.0-1254 & TXS0637-128          & 06:40:07.2 &  12:53:14.2 & bcu   & 0.49 & 24-02-2017 & 1500  &  1.5  & 17.9  \\
3FHLJ0702.6-1950 & TXS0700-197          & 07:02:42.9 & -19:51:22.0 & bll    & 0.42 & 10-03-2019 & 750   &  1.9  & 19.1  \\ 
3FHLJ0706.5+3744 & GB6J0706+3744        & 07:06:31.7 &  37:44:36.0 & bll    & 0.06 & 02-11-2018 & 1800  &  2.5  & 17.6  \\
3FHLJ0708.9+2240 & GB6J0708+2241        & 07:08:58.3 &  22:41:36.0 & bll    & 0.05 & 03-12-2018 & 4500  &  1.2  & 17.4  \\
3FHLJ0709.1-1525 & PKS0706-15           & 07:09:12.3 & -15:27:00.0 & bcu   & 0.55 & 04-12-2018 & 7200  &  2.5  & 18.9  \\
3FHLJ0723.0-0732 & 1RXSJ072259.5-073131 & 07:22:59.7 & -07:31:35.0 & bll    & 0.21 & 14-12-2018 & 3000  &  1.2  & 18.1  \\
3FHLJ0811.9+0237 & PMNJ0811+0237        & 08:12:01.8 &  02:37:33.0 & bll    & 0.02 & 04-12-2018 & 7200  &  2.0  & 18.0  \\
3FHLJ0816.4-1311 & PMNJ0816-1311        & 08:16:27.2 & -13:11:52.0 & bll    & 0.07 & 22-02-2018* & 5400 & 1.8 & 17.2 \\ 
3FHLJ0905.5+1357 & MG1J090534+1358      & 09:05:35.0 &  13:58:06.0 & bll    & 0.03 & 17-02-2018** & 850 & 1.2-2.0 & 16.2-17.3**  \\
3FHLJ0910.5+3329 & Ton1015              & 09:10:37.0 &  33:29:24.0 & bll    & 0.02 & 22-02-2018  & 350  & 1.5   & 16.2  \\
3FHLJ0953.0-0840 & PMNJ0953-0840        & 09:53:02.7 & -08:40:18.0 & bll    & 0.04 & 22-02-2018  & 900  & 1.5   & 16.8  \\
3FHLJ1037.6+5711 & GB6J1037+5711        & 10:37:44.3 &  57:11:56.0 & bll    & 0.01 & 30-04-2018  & 1050 & 1.2   & 16.5 \\
3FHLJ1055.6-0125 & NVSSJ105534-012617   & 10:55:34.3 & -01:26:16.0 & bll    & 0.04 & 14-12-2018  & 3000 & 1.3   & 18.6 \\
3FHLJ1059.1-1134 & PKSB1056-113         & 10:59:12.4 & -11:34:23.0 & bll    & 0.02  & 22-01-2019  & 900  & 1.8   & 17.6 \\
3FHLJ1150.5+4154 & RBS1040              & 11:50:34.7 &  41:54:40.9 & bll    & 0.02 & 20-01-2019  & 900  & 1.3   & 17.0  \\
3FHLJ1233.7-0145 & NVSSJ123341-014426   & 12:33:41.3 & -01:44:24.0 & bll    & 0.03 & 20-01-2019  & 3600 & 1.9   & 20.3 \\
3FHLJ1253.1+5300 & S41250+53            & 12:53:11.9 &  53:01:12.0 & bll    & 0.01 & 20-01-2019  & 1200 & 1.2   & 16.6 \\
3FHLJ1418.4-0233 & NVSSJ141826-023336   & 14:18:26.3 & -02:33:34.0 & bll    & 0.05 & 30-04-2018  & 1050 & 1.2   & 16.4  \\
3FHLJ1445.0-0326 & RBS1424              & 14:45:06.3 & -03:26:12.0 & bll    & 0.08 & 09-04-2019  & 3600 & 1.2   & 17.8 \\
3FHLJ1447.9+3608 & RBS1432              & 14:48:00.6 &  36:08:32.0 & bll    & 0.01 & 16-04-2019  & 1500 & 1.8   & 16.2 \\
3FHLJ1454.5+5124 & TXS1452+516          & 14:54:27.1 &  51:24:34.0 & bll    & 0.02 & 10-04-2019  & 3600 & 1.0   & 16.7 \\
3FHLJ1503.7-1541 & RBS1457              & 15:03:40.6 & -15:41:14.0 & bll    & 0.10 & 16-04-2019  & 3600 & 2.0   & 17.8 \\
3FHLJ1549.9-0659 & NVSSJ154952-065907   & 15:49:52.0 & -06:59:07.0 & bcu   & 0.14 & 09-04-2018  & 7200 & 1.7   & 18.1 \\
3FHLJ1748.6+7006 & S41749+70            & 17:48:32.8 &  70:05:50.7 & bll    & 0.03 & 29-03-2018  & 1500 & 0.9   & 16.6 \\    
3FHLJ1800.5+7827 & S51803+784           & 18:00:45.7 &  78:28:04.0 & bll    & 0.04 & 11-08-2018  & 900  & 2.5   & 16.7  \\
3FHLJ1841.3+2909 & MG3J184126+2910      & 18:41:21.7 &  29:09:41.0 & bcu    & 0.21 & 31-03-2018  & 1200 & 1.8   & 18.2 \\
3FHLJ1904.1+3627 & MG2J190411+3627      & 19:04:11.9 &  36:26:59.0 & bcu  & 0.08 & 07-04-2018  & 1600 & 2.0   & 17.3 \\  
3FHLJ1911.5-1908 & PMNJ1911-1908        & 19:11:29.7 & -19:08:23.0 & bcu  & 0.14 & 12-05-2018  & 1200 & 2.0   & 18.0  \\
3FHLJ1921.8-1607 & PMNJ1921-1607        & 19:21:51.5 & -16:07:13.2 & bll    & 0.16 & 18-10-2018*** & 5400 & 2.5 & 17.6 \\ 
\hline
\end{tabular}
\end{center}
\raggedright
\footnotesize \texttt{Col.1}: Fermi name of the target; \texttt{Col.2}: Counterpart name of the target; \texttt{Col.3 - 4}: Right ascension and declination of the optical counterpart; \texttt{Col.5}: Source classification reported in the 3FHL catalog (bll~=~BL Lac object, bcu~=~blazar candidate of uncertain type); \texttt{Col.6}: $E(B-V)$ taken from the NASA/IPAC Infrared Science Archive (https://irsa.ipac.caltech.edu/applications/DUST/); \texttt{Col.7}: Date of observation;  \texttt{Col.8}: Total integration time; \texttt{Col.9}: Seeing during the observation; \texttt{Col.10}: g magnitude measured from the acquisition image.\\ 
(*) This source was observed also in 14-12-2018 and 04-04-2019, (**) This source was observed also in 09-03-2019 and the source was found in two different flux states, (***) This source was also discussed in \citet{paiano2017ufo1} 
\end{table*}


\setcounter{table}{1}
\begin{table*}
\begin{center}
\caption{PROPERTIES OF THE OPTICAL SPECTRA OF THE 3FHL SAMPLE }\label{tab:results} 
\begin{tabular}{llcll} 
\hline
OBJECT           &   SNR       &   EW$_{min}$            &   z &  Line type  \\
\hline                                                                    
3FHLJ0009.4+5030 & 160 & 0.17 - 0.20 & ($>$~0.60)    &  L  \\
3FHLJ0015.7+5551 & 120 & 0.25 - 0.40 & 0.2168        &  E,G\\ 
3FHLJ0045.3+2127 & 90  & 0.30        & 0.4253        &  G   \\ 
3FHLJ0045.7+1217 & 200 & 0.20        & 0.2549        &  G   \\ 
3FHLJ0131.1+6120 & 75  & 0.20 - 0.70 & ($>$~0.10)    &  L  \\
3FHLJ0134.4+2638 & 80  & 0.35 - 0.50 & ($>$~0.15)    &  L  \\
3FHLJ0137.9+5815 & 145 & 0.25 - 0.30 & 0.2745        &  G   \\ 
3FGLJ0141.4-0929 & 200 & 0.20        & $\geq$~0.501   &  I   \\ 
3FHLJ0148.2+5201 & 110 & 0.30 - 0.35 & 0.437         &  G    \\ 
3FHLJ0241.3+6543 & 25  & 1.10 - 2.70 & 0.1211        &  E,G \\ 
3FHLJ0250.5+1712 & 100 & 0.35 - 0.45 & 0.2435        &  G   \\ 
3FHLJ0322.0+2336 & 160 & 0.10 - 0.25 & ($>$~0.25)    &  L  \\
3FHLJ0423.8+4149 & 40  & 0.35 - 1.60 & 0.3977        &  E   \\ 
3FHLJ0433.1+3227 & 110 & 0.25 - 0.35 & ($>$~0.45)    &  L  \\ 
3FHLJ0433.6+2905 & 15  & 1.35 - 3.80 & 0.91~?        &  E   \\ 
3FHLJ0434.7+0921 & 75  & 0.30 - 0.90 & ($>$~0.1)     &  L  \\
3FHLJ0500.3+5238 & 40  & 0.55 - 4.00 & 0.1229        &  E,G \\
3FHLJ0506.0+6113 & 70  & 0.30 - 0.90 & 0.538~?        &  G   \\ 
3FHLJ0515.8+1528 & 130 & 0.20 - 0.30 & ($>$~0.20)    &  L  \\
3FHLJ0540.5+5823 & 60  & 0.40 - 0.90 & ($>$~0.10)    &  L  \\
3FHLJ0600.3+1245 & 50  & 0.45 - 1.85 & 0.0835        &  E,G \\ 
3FHLJ0601.0+3837 & 40  & 0.50 - 1.35 & 0.662~?       &  G   \\
3FHLJ0602.0+5316 & 70  & 0.35 - 1.00 & 0.0522        &  G   \\
3FHLJ0607.4+4739 & 100 & 0.15 - 0.40 & ($>$~0.10)    &  L  \\
3FHLJ0612.8+4122 & 50  & 0.35 - 1.10 & $\geq$1.107   &  I   \\ 
3FHLJ0620.6+2645 & 75  & 0.40 - 0.80 & 0.1329        &  G   \\ 
3FHLJ0640.0-1254 & 80  & 0.35 - 0.65 & 0.1365        &  E,G \\
3FHLJ0702.6-1950 & 50  & 0.50 - 0.95 & ($>$~0.10)    &  L   \\
3FHLJ0706.5+3744 & 100 & 0.20 - 0.35 & $\geq$0.1042  &  I   \\ 
3FHLJ0708.9+2240 & 250 & 0.15 - 0.20 & 0.2966        &  G   \\ 
3FHLJ0709.1-1525 & 70  & 0.35 - 1.10 & 0.1420        &  E,G \\ 
3FHLJ0723.0-0732 & 170 & 0.10 - 0.25 & 0.3285        &  G   \\ 
3FHLJ0811.9+0237 & 70  & 0.40 - 0.85 & 0.1726        &  E,G \\  
3FHLJ0816.4-1311 & 300 & 0.10        & ($>$~0.40)    &  L  \\
3FHLJ0905.5+1357 & 200 & 0.10 - 0.20 & 0.2239:       &  E   \\ 
3FHLJ0910.5+3329 & 200 & 0.10 - 0.20 & ($>$~0.15)   &  L  \\
3FHLJ0953.0-0840 & 220 & 0.15 - 0.25 & ($>$~0.15)   &  L  \\
3FHLJ1037.6+5711 & 300 & 0.10 - 0.15 & ($>$~0.25)   &  L  \\
3FHLJ1055.6-0125 & 130 & 0.20 - 0.40 & ($>$~0.55)   &  L  \\
3FHLJ1059.1-1134 & 50  & 0.30 - 1.15 & ($>$~0.10)   &  L  \\
3FHLJ1150.5+4154 & 200 & 0.15 - 0.25 & ($>$~0.25)   &  L  \\
3FHLJ1233.7-0145 & 40  & 0.45 - 0.95 & ($>$~0.10)   &  L    \\
3FHLJ1253.1+5300 & 250 & 0.10 - 0.20 & $\geq$0.6638 &  I   \\ 
3FHLJ1418.4-0233 & 200 & 0.15 - 0.30 & ($>$~0.12)   &  L  \\
3FHLJ1445.0-0326 & 200 & 0.10 - 0.20 & ($>$~0.45)   &  L  \\
3FHLJ1447.9+3608 & 250 & 0.10 - 0.20 & $\geq$0.738  &  I   \\ 
3FHLJ1454.5+5124 & 320 & 0.05 - 0.10 & ($>$~0.40)   &  L  \\
3FHLJ1503.7-1541 & 50  & 0.55 - 1.25 & ($>$~0.10)   &  L  \\
3FHLJ1549.9-0659 & 200 & 0.15 - 0.20 & 0.418        &  G   \\ 
3FHLJ1748.6+7006 & 320 & 0.10 - 0.20 & ($>$~0.3)    &  L    \\
3FHLJ1800.5+7827 & 130 & 0.25 - 0.40 & 0.683        &  E     \\ 
3FHLJ1841.3+2909 & 40  & 0.55 - 1.35 & ($>$~0.10)   &  L    \\
3FHLJ1904.1+3627 & 60  & 0.40 - 1.50 & 0.08977       &  G     \\ 
3FHLJ1911.5-1908 & 50  & 0.55 - 1.20 & 0.138        &  E,G   \\
3FHLJ1921.8-1607 & 125 & 0.20 - 0.50 & ($>$~0.12)  &  L    \\
\hline
\end{tabular}
\end{center}
\raggedright
\footnotesize 
\texttt{Col.1}: Name of the target; \texttt{Col.2}: Median S/N of the spectrum; \texttt{Col.3}: Range of the minimum equivalent width (EW$_{min}$) derived from different regions of the spectrum; \texttt{Col.4}: Redshift; \texttt{Col.5}: Type of detected line to estimate the redshift: \textit{E} = emission line, \textit{G} = galaxy absorption line, \textit{I}= intervening absorption assuming Mg~II 2800$\textrm{\AA}$ identification, \textit{L}= lower limit derived on the lack of detection of host galaxy absorption lines assuming a BLL elliptical host galaxy with M(R) = -22.9 \citep[see details in][]{paiano2017tev}.\\
(:) This marker indicates that the redshift is tentative due to the detection of only one feature. \\
(*) For this source we found other two absorption line systems due to Fe~II~(2382, 2600) (See details text).  
\end{table*}

\newpage
\setcounter{table}{2}
\begin{table}
\centering
\caption{MEASUREMENTS OF THE DETECTED LINES}\label{tab:ew}
\begin{center}
\begin{tabular}{lclll} 
\hline
Object name          &  $\lambda$   &    EW   &     Line ID   & Line Lum.\\

                              &     ($\textrm{\AA}$)         &    ($\textrm{\AA}$)           &                    &  (erg/s)    \\
\hline
3FHLJ0015.7+5551 & 4787  & 1.6 & Ca II 3934    \\
                 & 4829  & 1.0 & Ca II 3968    \\
                 & 5238  & 1.0 & G-band 4305   \\
                 & 6093  & -0.5 & [O III] 5007 &  4.2$\times$10$^{40}$ \\
                 & 6411  & 0.4 & Ca+Fe   5269  \\
3FHLJ0045.3+2127 & 5607  & 1.0 & Ca II 3934    \\
                 & 5656  & 0.5 & Ca II 3968      \\
3FHLJ0045.7+1217 & 4679  & -0.2 & [O II] 3727 &  2.3$\times$10$^{40}$   \\
                 & 4936  & 1.0 & Ca II 3934    \\
                 & 4980  & 0.9 & Ca II 3968     \\
                 & 5403  & 0.3 & G-band 4305   \\
                 & 6489  & 0.6 & Mg~I 5175      \\
                 & 6612  & 0.5 & Ca+Fe  5269   \\
                 & 7393  & 0.5 & Na~I 5892      \\
3FHLJ0137.9+5815 & 5013 & 0.4 & Ca~II 3934   \\
                 & 5057 & 0.6 & Ca~II 3968    \\
                 & 6596 & 0.9 & Mg~I 5175     \\
3FGLJ0141.4-0929 & 4197 & 1.0  & Mg~II 2796    \\
                 & 4207 & 0.5  & Mg~II 2803    \\
3FHLJ0148.2+5201 & 5654 & 1.3  & Ca~II 3934     \\
                 & 5703 & 1.5  & Ca~II 3968    \\
                 & 6183 & 0.8  & G-band 4305   \\
3FHLJ0241.3+6543 & 6607 & 2.8  & Na~I 5892            \\
                 & 7358 & -1.7  & H$_{\alpha}$~6563 &  1.3$\times$10$^{41}$    \\
                 & 7381 & -2.5  & [N~II]~6584   &  2.0$\times$10$^{41}$       \\
                 & 7531 & -2.0  & [SII]~6716   &  1.3$\times$10$^{41}$        \\
                 & 7545 & -1.4  & [SII]~6731  &  8.7$\times$10$^{40}$        \\
3FHLJ0250.5+1712 & 4890 & 1.7  & Ca~II 3934          \\
                 & 4932 & 1.5  & Ca~II 3968          \\
                 & 5090 & 0.5  & H$_{\delta}$~4102    \\
                 & 5352 & 0.6  & G-band 4305          \\
                 & 6045 & 0.6  & H$_{\beta}$~4861     \\
                 & 6433 & 0.8  & Mg~I 5175            \\
                 & 6549 & 0.6  & Ca+Fe 5269           \\
                 & 7324 & 1.9  & Na~I 5892            \\
3FHLJ0423.8+4149 & 6998 & -1.9 & [O III] 5007 &  5.5$\times$10$^{41}$       \\
3FHLJ0433.6+2905 & $\sim$5340 & -9.5  & Mg~II 2800  & 2.5$\times$10$^{42}$      \\
3FHLJ0500.3+5238 & 4833 & 3.5 & G-band 4305         \\
                 & 5459 & 3.1 & H$_{\beta}$ 4861    \\
                 & 5811 & 1.6 & Mg~I~5175           \\
                 & 5917 & 2.3 & Ca+Fe~5269          \\
                 & 6617 & 4.7 & Na~I~5892           \\
                 & 7369 & -0.3 & H$_{\alpha}$~6563 & 1.7$\times$10$^{40}$      \\
                 & 7393 & -1.3 & [N~II]~6584   & 8.7$\times$10$^{40}$     \\
3FHLJ0506.0+6113 & 6042 & 0.8 & Ca~II 3934   \\
                 & 6096 & 0.6 & Ca~II 3968   \\
3FHLJ0600.3+1245 & 4664 & 1.6  & G-band 4305         \\
                 & 5267 & 1.8  & H$_{\beta}$ 4861    \\
                 & 5608 & 2.3  & Mg~I~5175          \\
                 & 5709 & 1.6  & Ca+Fe~5269          \\
                 & 6385 & 3.1  & Na~I~5892           \\
                 & 7111 & -1.2 & H$_{\alpha}$~6563 & 2.3$\times$10$^{40}$         \\
                 & 7133 & -1.8 & [N~II]~6584  & 3.5$\times$10$^{40}$        \\
3FHLJ0601.0+3837 & 6538 & 2.5 & Ca~II 3934     \\
                 & 6596 & 2.0 & Ca~II 3968     \\
                 &  &  & &   \\
\hline
\end{tabular}
\end{center}
\raggedright
\footnotesize  \texttt{Col.1}: Name of the target; \texttt{Col.2}: Barycenter of the detected line; \texttt{Col.3}: Measured equivalent width; \texttt{Col.4}: Line identification; \texttt{Col.5}: Line luminosity.\\
\end{table}

\setcounter{table}{2}
\begin{table}
\caption{MEASUREMENTS OF THE DETECTED LINES} 
\centering
\begin{center}
\begin{tabular}{lclll} 
\hline
Object name          &  $\lambda$   &    EW   &     Line ID   & Line Lum.\\

                              &      ($\textrm{\AA}$)         &    ($\textrm{\AA}$)           &                    &  (erg/s)    \\
\hline
3FHLJ0602.0+5316 & 4139 & 2.8 & Ca~II 3934          \\
                 & 4176 & 1.5 & Ca~II 3968          \\
                 & 4529 & 2.1 & G-band 4305         \\
                 & 5115 & 0.6 & H$_{\beta}$ 4861    \\
                 & 5446 & 1.3 & Mg~I~5175           \\
                 & 5543 & 0.7 & Ca+Fe~5269          \\
                 & 6200 & 1.8 & Na~I~5892           \\
3FHLJ0612.8+4122 & 5022 & 1.5 & Fe~II~2382    \\
                 & 5450 & 0.4 & Fe~II~2586    \\
                 & 5478 & 1.5 & Fe~II~2600    \\
                 & 5891 & 3.5 & Mg~II 2796    \\
                 & 5906 & 3.1 & Mg~II 2803    \\
3FHLJ0620.6+2645 & 4456 & 2.8 & Ca~II 3934     \\
                 & 4496 & 3.7 & Ca~II 3968     \\
                 & 4876 & 1.7 & G-band 4305          \\
                 & 5507 & 1.2 & H$_{\beta}$ 4861     \\
                 & 5863 & 2.6 & Mg~I~5175             \\
                 & 5969 & 0.8 & Ca+Fe~5269            \\
                 & 6676 & 2.1 & Na~I~5892             \\
                 & 7435 & 0.4 & H$_{\alpha}$~6563     \\
3FHLJ0640.0-1254 & 4471 & 0.4  & Ca~II 3934 \\
                 & 4510 & 0.8  & Ca~II 3968          \\
                 & 4892 & 1.6  & G-band 4305         \\
                 & 5525 & 0.7  & H$_{\beta}$ 4861    \\
                 & 5882 & 2.2* & Mg~I~5175           \\
                 & 5988 & 0.7  & Ca+Fe~5269          \\
                 & 6697 & 1.4  & Na~I~5892           \\
                 & 7482 & -0.6 & [N~II]~6584  & 4.7$\times$10$^{40}$        \\
3FHLJ0706.5+3744 & 4344 & 0.4 & Ca~II 3934           \\
                 & 4382 & 0.6 & Ca~II 3968           \\
3FHLJ0708.9+2240 & 5100 & 0.4 & Ca~II 3934     \\
                 & 5146 & 0.5 & Ca~II 3968     \\
                 & 5581 & 0.3 & G-band 4305    \\
                 & 6710 & 0.5 & Mg~I~5175      \\
                 & 6832 & 0.2 & Ca+Fe~5269     \\
3FHLJ0709.1-1525 & 4492 & 5.3  & Ca~II 3934           \\
                 & 4532 & 4.6  & Ca~II 3968           \\
                 & 4916 & 1.7  & G-band 4305          \\
                 & 5552 & 1.9  & H$_{\beta}$ 4861     \\
                 & 6017 & 1.3  & Ca+Fe~5269           \\
                 & 6729 & 2.3  & Na~I~5892            \\
                 & 7518 & -1.0 & [N~II]~6584 & 3.4$\times$10$^{40}$           \\
3FHLJ0723.0-0732 & 5226 & 0.5 & Ca~II 3934     \\
                 & 5272 & 0.9 & Ca~II 3968     \\
                 & 5718 & 0.5 & G-band 4305    \\
3FHLJ0811.9+0237 & 4613 & 4.7 & Ca~II 3934          \\
                 & 4653 & 4.5 & Ca~II 3968          \\
                 & 5047 & 1.6 & G-band 4305         \\
                 & 5700 & 2.7 & H$_{\beta}$~4861    \\
                 & 6069 & 3.0 & Mg~I~5175           \\
                 & 6178 & 1.9 & Ca+Fe~5269          \\
                 & 6910 & 6.0* & Na~I~5892          \\
                 & 7720 & -0.3 & [N~II]~6584  & 9.8$\times$10$^{39}$        \\
3FHLJ0905.5+1357 & 6128 & -0.7** & [O III]~5007  & 8$\times$10$^{40}$**   \\
3FHLJ1253.1+5300 & 4652 & 0.3 & Mg~II 2796    \\
                 & 4664 & 0.2 & Mg~II 2803   \\
%
                 \hline
\end{tabular}
\end{center}
\raggedright
\footnotesize  (*) This marker indicates that the line is partially contaminated by a telluric band or the interstellar Na~I absorption of our Galaxy,\\
(**) This is the average EW (EW~=~0.5~$\textrm{\AA}$ when the source is in high state and EW~=~0.9~$\textrm{\AA}$ in the low state).
\end{table}

\setcounter{table}{2}
\begin{table}
\caption{MEASUREMENTS OF THE DETECTED LINES} 
\centering
\begin{center}
\begin{tabular}{lcllll} 
\hline
Object name          &  $\lambda$   &    EW   &     Line ID   & Line Lum.\\

                              &     ($\textrm{\AA}$)         &    ($\textrm{\AA}$)           &                    &  (erg/s)    \\
\hline
3FHLJ1447.9+3608 & 4859 & 0.4 & Mg~II 2796   \\
                 & 4872 & 0.2 & Mg~II 2803    \\
3FHLJ1549.9-0659 & 5578 & 1.1 & Ca~II 3934     \\
                 & 5627 & 0.9 & Ca~II 3968    \\
                 & 6104 & 0.5 & G-band 4305    \\
3FHLJ1800.5+7827 & 4712 & -8.3 & Mg~II 2800  & 2.2$\times$10$^{43}$      \\
3FHLJ1904.1+3627 & 4287 & 6.0 & Ca~II 3934          \\ 
                 & 4325 & 5.9 & Ca~II 3968          \\ 
                 & 4691 & 3.6 & G-band 4305         \\ 
                 & 5298 & 2.6 & H$_{\beta}$ 4861    \\ 
                 & 5640 & 3.5 & Mg~I~5175           \\ 
                 & 5742 & 2.1 & Ca+Fe~5269          \\ 
                 & 6421 & 3.6 & Na~I~5892           \\ 
3FHLJ1911.5-1908 & 4242 & -2.8 & [O~II]~3727 & 8.3$\times$10$^{40}$\\
                 & 4474 & 4.1 & Ca~II 3934 \\
                 & 4514 & 2.9 & Ca~II 3968 \\
                 & 4896 & 3.3 & G-band 4305 \\
                 & 5529 & 1.3 & H$_{\beta}$ 4861 \\
                 & 5695 & -0.9 & [O~III]~5007 & 2.8$\times$10$^{40}$\\
                 & 5886 & 3.3 & Mg~I~5175 \\
                 & 5993 & 2.2 & Ca+Fe~5269 \\
                 & 6702 & 3.0 & Na~I~5892 \\
                 & 7448 & -0.7* & [N~II]~6548 & 1.5$\times$10$^{40}$\\
                 & 7465 & -0.6* & H$_{\alpha}$~6563 & 1.3$\times$10$^{43}$\\
                 & 7489 & -2.6 & [N~II]~6584 & 6.3$\times$10$^{40}$\\
\hline
\end{tabular}
\end{center}
\raggedright
\footnotesize (*) These lines are very faint and blended.
\end{table}


\setcounter{figure}{0}
\begin{figure*}
\includegraphics[width=1.35\textwidth, angle=-90]{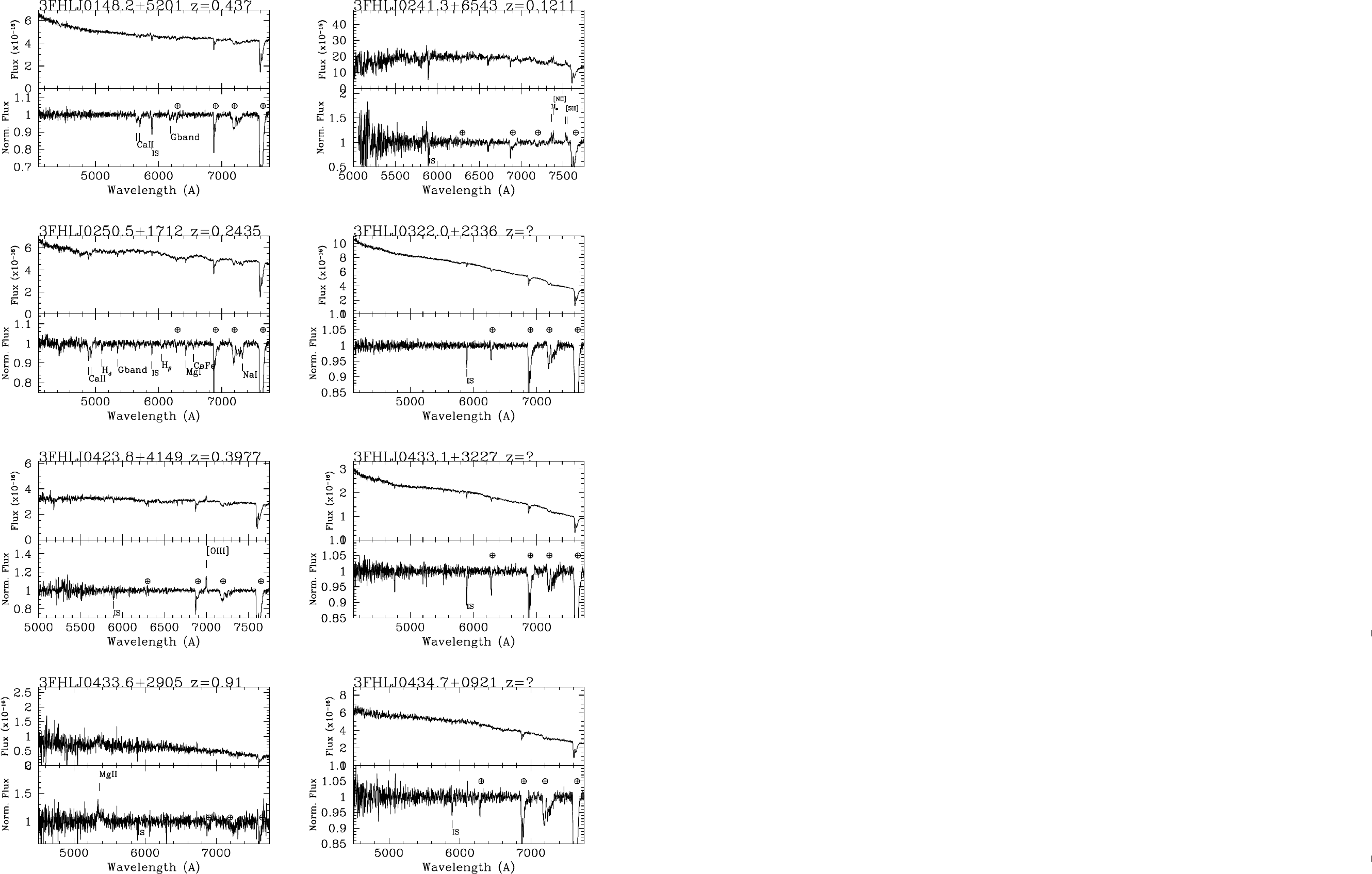}
\caption{- Continued -}
\end{figure*}

\setcounter{figure}{0}
\begin{figure*}
\includegraphics[width=1.35\textwidth, angle=-90]{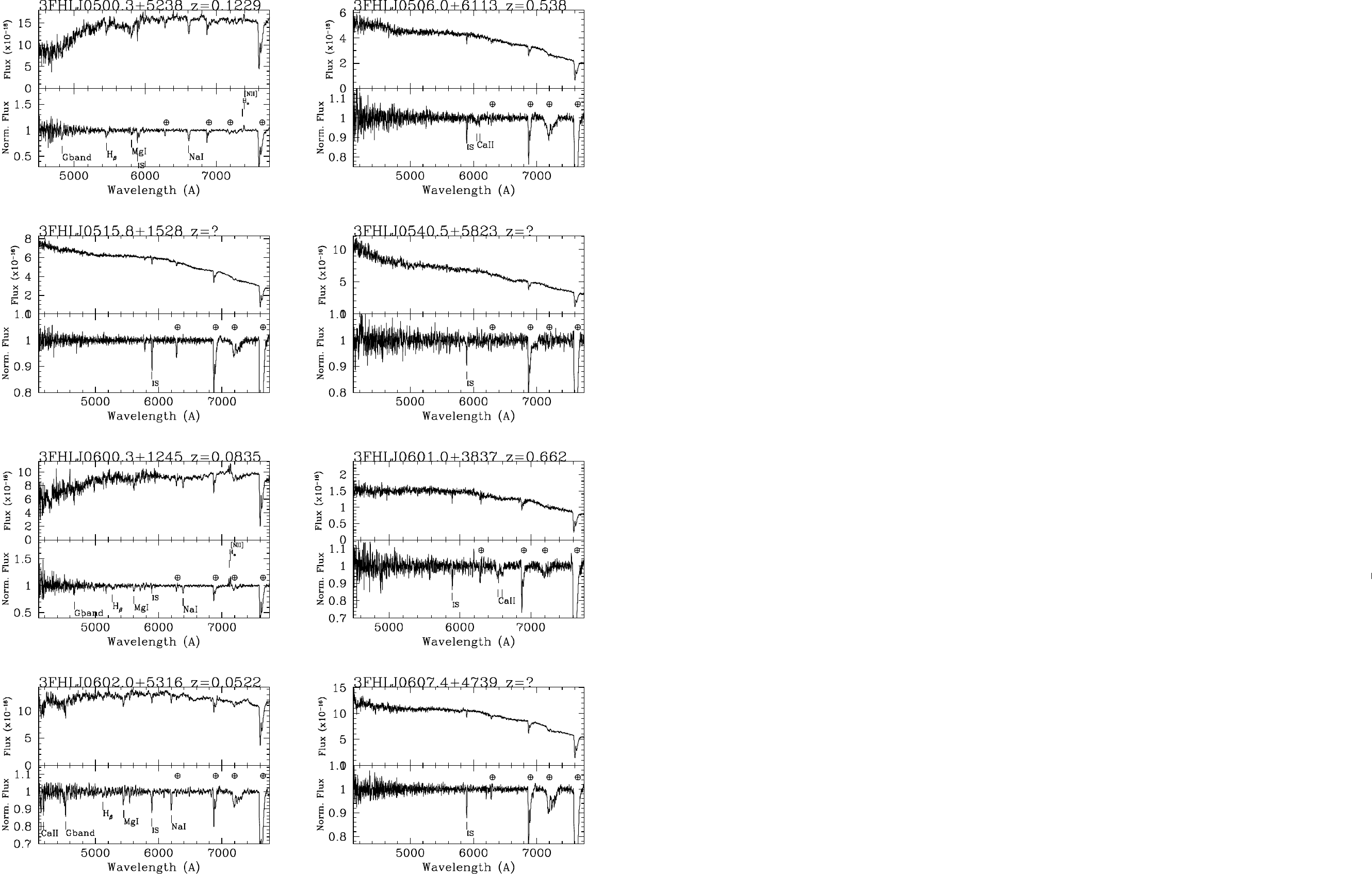}
\caption{- Continued - }
\end{figure*}

\setcounter{figure}{0}
\begin{figure*}
\includegraphics[width=1.35\textwidth, angle=-90]{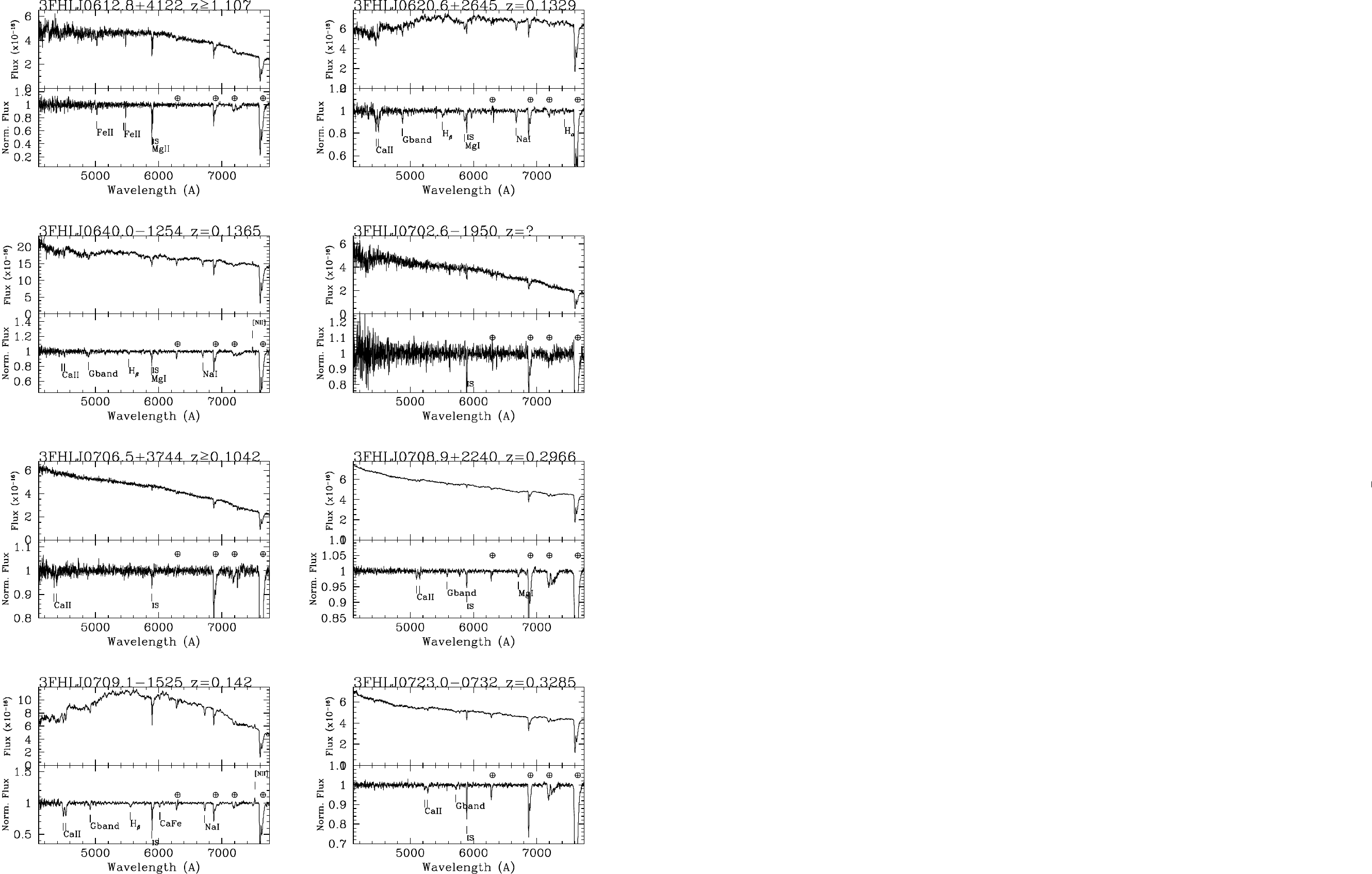}
\caption{- Continued -}
\end{figure*}

\setcounter{figure}{0}
\begin{figure*}
\includegraphics[width=1.35\textwidth, angle=-90]{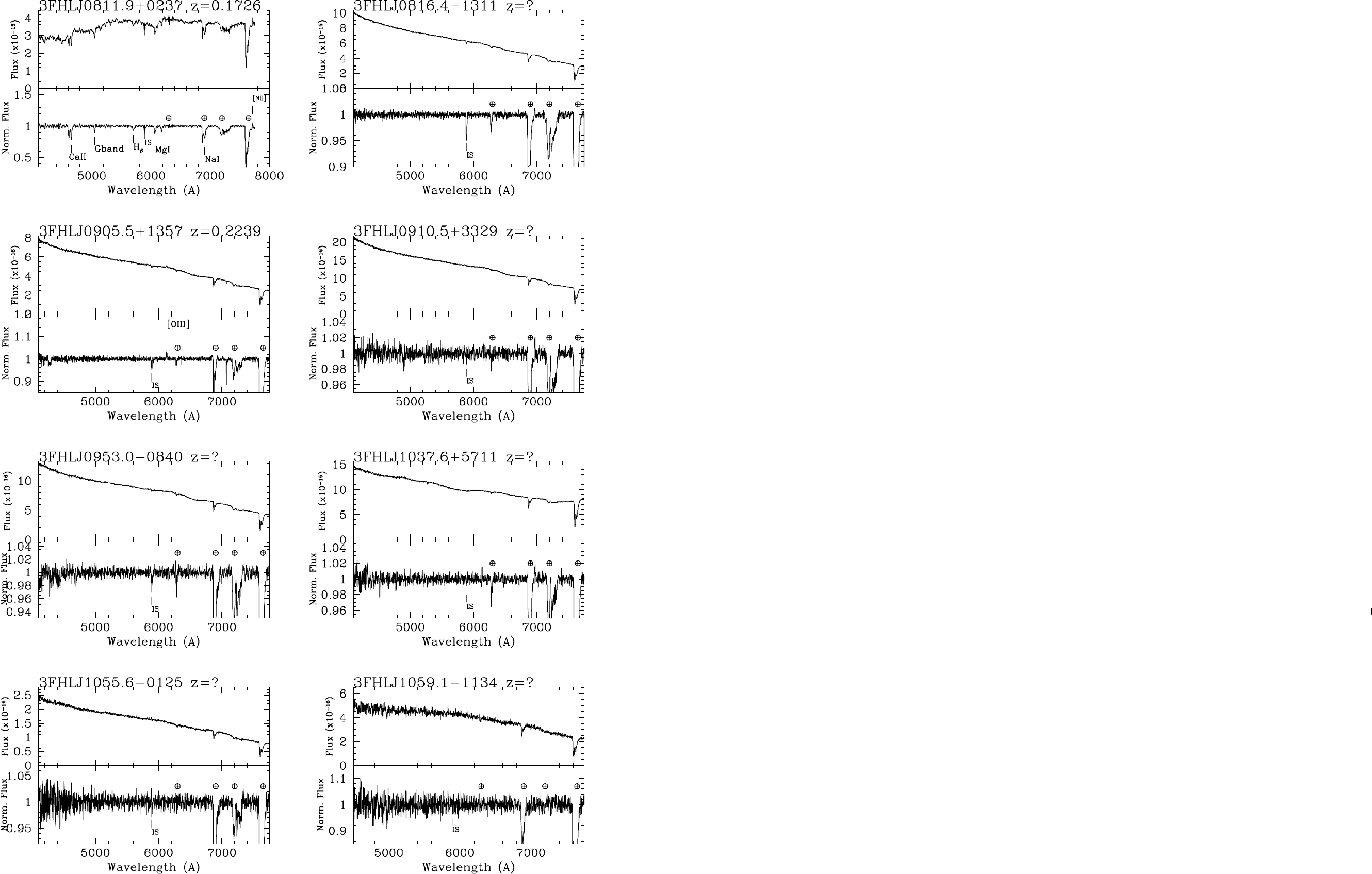}
\caption{- Continued -}
\end{figure*}

\setcounter{figure}{0}
\begin{figure*}
\includegraphics[width=1.35\textwidth, angle=-90]{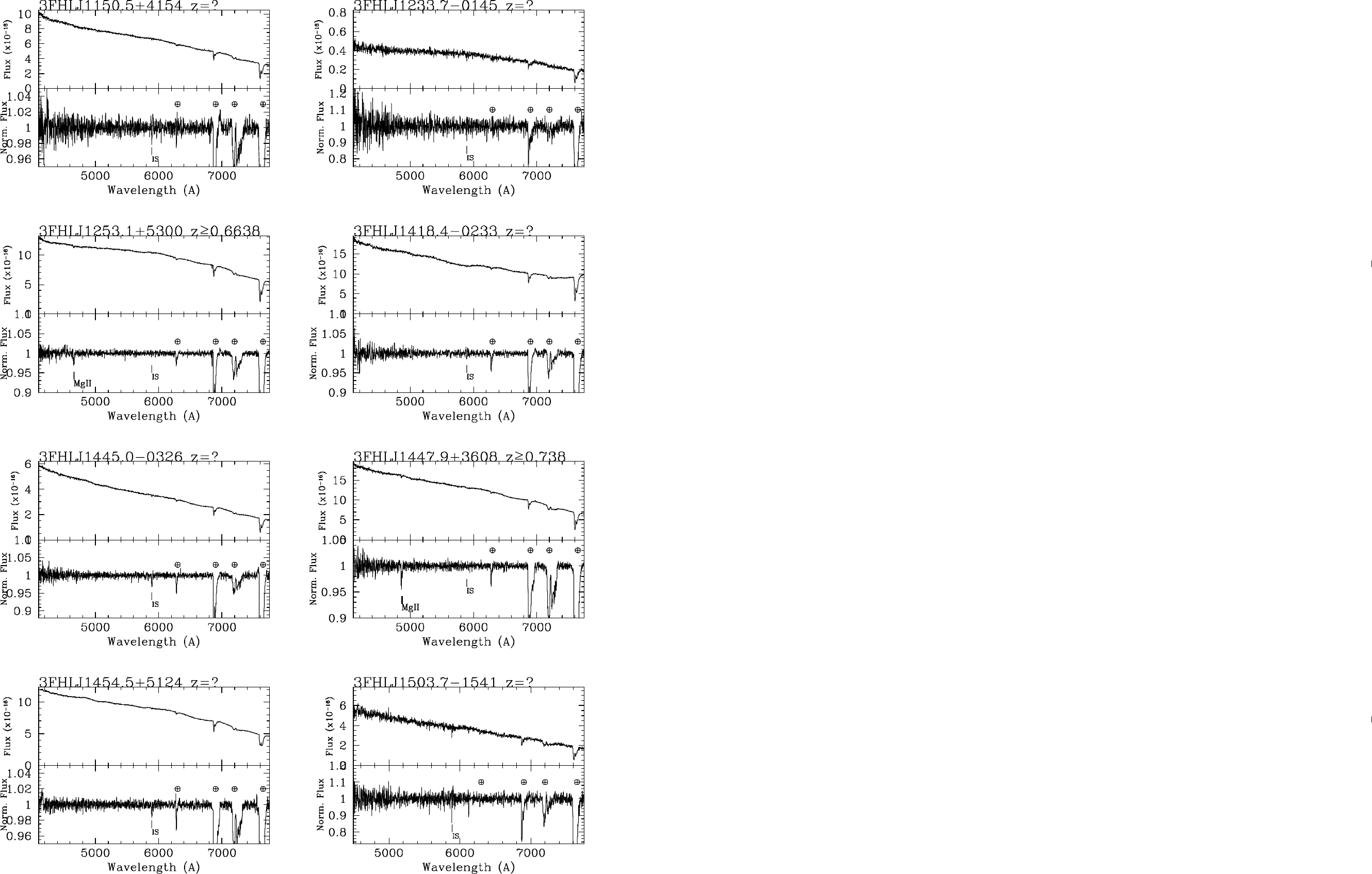}
\caption{- Continued -}
\end{figure*}

\setcounter{figure}{0}
\begin{figure*}
\includegraphics[width=1.35\textwidth, angle=-90]{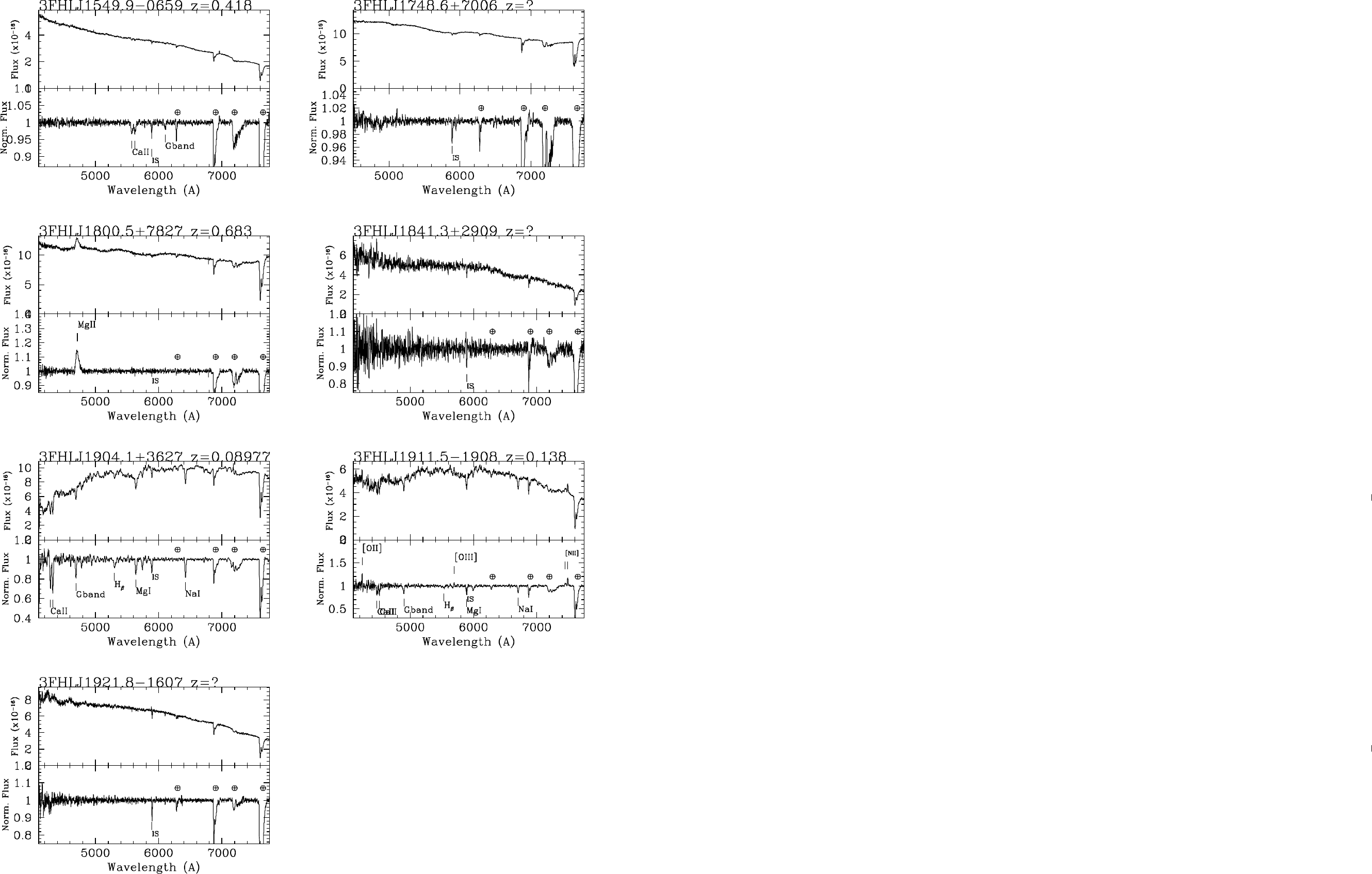}
\caption{- Continued - }
\end{figure*}


\section*{Data availability}
The flux calibrated and dereddened spectra are available in our online database ZBLLAC\footnote{http://web.oapd.inaf.it/zbllac/}. 



\bsp	

\end{document}